\def\year{2021}\relax
\documentclass[letterpaper]{article} 
\usepackage{aaai21}  
\usepackage{times}  
\usepackage{helvet} 
\usepackage{courier}  
\usepackage[hyphens]{url}  
\usepackage{graphicx} 
\usepackage{natbib}  
\usepackage{caption} 
\usepackage{booktabs}
\usepackage{multirow}
\usepackage[caption = false]{subfig}
\usepackage{changes}
\nocopyright

\title{Comparing the Language of QAnon-related content on Parler, Gab, and Twitter}
\author {
    Andrea \v{S}ipka\footnote{Corresponding author, email: sipka at ifi dot uzh dot ch}, Anik\'{o} Hann\'{a}k, Aleksandra Urman \\
}
\affiliations {
    Social Computing Group, University of Zurich
}

\begin{document}

\maketitle

\begin{abstract}
Parler, a ``free speech'' platform popular with conservatives, was taken offline in January 2021 due to the lack of moderation of hateful and QAnon- and other conspiracy-related content that allegedly made it instrumental in the organisation of the storming of the US Capitol on January 6. However, Parler co-existed with other social media platforms, and comparative studies are needed to draw conclusions about the prevalence of anti-social language, hate speech, or conspiracy theory content on the platform. We address this through a cross-platform comparison of posts related to QAnon. We compare posts with the hashtag \#QAnon on Parler over a month-long period with posts on Twitter and Gab.
In our analysis, Parler emerges as the platform with the highest average toxicity of the posts, though this largely stems from the distinctive way hashtags are used on this platform. Gab has the highest proportion of \#QAnon posts with hate terms, and Parler and Twitter are similar in this respect. On all three platforms, posts mentioning female political figures, Democrats, or Donald Trump have more anti-social language than posts mentioning male politicians, Republicans, or Joe Biden. An analysis of entities mentioned in posts revealed differences in content - with Twitter posts mentioning prominent figures related to QAnon, while Parler and Gab posts mention entities related to conspiracies. Narrative analysis indicates that the discussion on Twitter centres on QAnon and is critical of it, Parler focuses on supporting Donald Trump, while on Gab the discussion focuses on more conspiratorial content, in relation to both Trump and other political figures.
\end{abstract}

\section{Introduction}

On January 6, 2021, a group of supporters of the President of the United States Donald Trump stormed the US Capitol building, interrupting a joint congressional session that was counting the electoral votes following the November 2020 presidential election and was due to confirm Joe Biden as the winner. For two months prior to the storming, Donald Trump repeatedly claimed that the election was being stolen. Trump's supporters started movements on multiple social networks (e.g. \#StopTheSteal) and held protests across the United States, some of which turned violent. The culmination of that movement, the Capitol riot itself, resulted in five deaths, with more than 140 injured, and was the subject of the second impeachment trial of Donald Trump.

Following the riot, various media outlets made attempts to uncover how the riot had been organised. Many reputable media outlets identified Parler, a social networking site launched in 2018 as a ``free speech alternative'' to more established platforms such as Twitter, as the place where the organisation took place \cite{frenkel2021, timberg2021}. In connection to these allegations, in the days following the riot, a growing number of service providers pulled support from Parler, and it was removed from the most popular app stores. This ultimately resulted in Parler being taken offline on January 10, 2021, due to the inability to host the service. Parler returned online on February 15, 2021, following the platform's migration to a set of alternative service providers, although with all the previous data removed. Since its inception, Parler was increasingly popular with users on the right of the political spectrum, with many conservative commentators and politicians joining and advocating for the platform between 2018 and 2020.

The 8th most popular hashtag on Parler prior to the shutdown was \#QAnon \cite{aliapoulios2021early}, used to denote a far-right conspiracy theory suggesting that ``Trump has been battling against a satan worshipping global child sex-trafficking ring and an anonymous source called ‘Q’ is cryptically providing secret information about the ring'' \cite{Zuckerman2019QAnon}. QAnon-related hashtags were amongst the top hashtags used in profile descriptions of users spreading misinformation related to the 2020 presidential election on Twitter \cite{eip2020}, and among the top 10 hashtags in a sample of 600 million US election-related tweets \cite{ferrara2020}. The followers of QAnon have been spreading conspiracy theories relating to elections, and have participated in the storming of the Capitol (most notably, ``QAnon Shaman'' Jake Angeli) \cite{angeliCharge}. For this reason, we analyse posts containing the hashtag \#QAnon in the period between December 7, 2020, and January 10, 2021, as it encompasses multiple events of interest: the ``Stop the steal'' movement, the US Senate run-off election in Georgia, multiple protests of election results, including the January 6 riot, the confirmation of Biden as the election winner, Twitter banning Trump, and the announcement of the (temporary) shutdown of Parler.

Despite the attention Parler has garnered in relation to the January 6 riot, and the consequences the platform has suffered in connection to the claims about the insufficient moderation or prevention of spread of dangerous and egregious content, Parler was not the only social media platform used by the rioters, as evidenced by the court documents in cases brought against the participants \cite{brewster2021}. Parler exists in a large ecosystem of platforms, which have varied functionalities and moderation policies. While many researchers have studied political movements on social media platforms, few studies have been published on Parler. To make a fair judgment about the difference in discourse, tone, and anti-social language, the scientific community needs to conduct studies comparing multiple similar platforms, yet comparative studies are rare.

We aim to partially address the existing gap with the present study encompassing three platforms - Parler, Twitter, and Gab. We choose Twitter and Gab as comparison points to Parler for two reasons. Firstly, the three platforms are very similar in basic functionality, being primarily micro-blogging platforms with the same mechanism of posting and engaging with the content, which makes them comparable. We do not include other platforms with fundamentally different architectures (e.g. Facebook or Reddit) as that would decrease the internal validity of the analysis. Secondly, as insufficient content moderation was frequently cited as a reason for removing Parler's access to services, a comparison with platforms that have opposing types of moderation policies is particularly relevant - Twitter has been engaging in moderation of QAnon content \cite{twitter2020}, while Gab is marketing itself in similar ways Parler did (\textit{``A social network that champions free speech, individual liberty and the free flow of information online''}), and has been reported as one of the platforms that Parler users migrated to following the shutdown \cite{wilson2021}.

\subsection{Research Questions}
Our aim is to compare Parler, Twitter, and Gab posts that have used the same hashtag in the same time period, to gain an understanding of differences and similarities between the three platforms in the context of the discourse surrounding \#QAnon in the run-up, and for a few days following, the Capitol riot. We analyse the volume of posting and the number of users who post to gauge the activity and interest levels across three platforms related to QAnon. In addition, as lack of moderation of threatening and harmful content was a frequent criticism of Parler, we analyse multiple aspects of anti-social language (such as hateful, threatening, and toxic language). This allows us to draw conclusions on how Parler compares to similar moderated and unmoderated platforms. We look at named entities used across the platforms, focusing on similarities and differences, to increase understanding of which figures, events and similar entities are mentioned in posts. In particular, as the period analysed includes major political events in the US, we test if the language differs in posts mentioning groups of political figures. We look at the posts about politicians from the opposite political parties (as Gab and Parler are considered right-wing), and posts mentioning politicians of different genders\footnote{male and female only, due to a lack of posts mentioning gender non-binary politicians in the sample} (as previous research reported higher levels of incivility on social media towards women in politics \cite{rheault2019politicians}). Finally, we analyse the prevalent narratives on the three platforms to gain a deeper understanding of differences and similarities in the conversation. We formulate these aims as research questions:
\begin{itemize}
    \item \textbf{RQ1}: How does the volume of \#QAnon posts, and the number of users posting, vary across the three platforms?
    \item \textbf{RQ2}: How do the three platforms compare with respect to the prevalence of anti-social language?
    \item \textbf{RQ3}: Are there differences in the relative prevalence of themes and political figures mentioned between the three platforms, and in the use of anti-social language related to those mentions?
    \item \textbf{RQ4}: What are the prevalent narratives on the three platforms, and how do they compare?
\end{itemize}

\subsection{Ethical considerations}
For both Parler and Gab, we used unofficial APIs to collect the data. This data was, at the time of collection, publicly available to anyone with an account on the platform (Parler), or anyone on the internet (Gab). Our data collection was non-intrusive as it did not affect users or platforms. The collection of Twitter data is in line with the platform's Terms of use, through the official API. For all three platforms, while the original data included usernames, bios, and similar information which could lead to the identification of individuals, we have excluded these fields from the analysis, thus anonymising the data. The posts were only collected for accounts that were not private at the time.

\section{Related work}

\subsubsection{Twitter}
Twitter is one of the most, if not the most, studied social network. Regarding QAnon-related content in particular, a study of banned Twitter users found that ``QAnon is well-positioned at the centre of the political hashtag community'' of banned users \cite{chowdhury2020twitter}. Numerous studies have been conducted on hate speech on Twitter. For the present research, of particular relevance are studies of hate speech related to violent or political events. A study of hate speech and white nationalist language on Twitter around and after the 2016 US presidential election reveals ``no evidence of an increase in hate speech before or after the election'', while noting that there are short time periods where the level of hate rises, finding ``evidence of tens of thousands of tweets containing hate speech and white nationalist rhetoric on Twitter'' \cite{siegel16trumping}.
\subsubsection{Gab}
To the best of our knowledge, no studies examining QAnon on Gab have been conducted, although a study of topic evolution on the platform found that by the start of 2018, the discourse on Gab has switched to alt-right political topics, and in particular posts related to QAnon \cite{mcilroy2019welcome}. Related to language on the platform, previous research on Gab found that 5.4\% of Gab posts contain at least one hate word, and that the most prevalent points of discussion on Gab are news, events, and conspiracy theories \cite{zannettou2018gab}. 90\% of posts on Gab were found to have toxicity scores less than 0.7 (on a scale from 0 to 1), although the authors note that there are users who ``abuse the lack of moderation to spread hate'' \cite{lima2018inside}.
\subsubsection{Parler}
Most of the research on Parler is still undergoing peer review, and to the best of our knowledge, nobody examined QAnon or anti-social language on the platform. However, studies conducted include reports of the number of users on Parler having more than doubled within weeks around the 2020 US election, and that Parler ``has emerged as a space in which accounts that have been suspended by Twitter Safety continue to communicate with their audiences'' \cite{thiel2021contours}. A study supplementary to a release of a large Parler dataset has found that Parler has experienced growth in user base in close proximity to ``online censorship on mainstream platforms like Twitter, as well as events related to US politics'' \cite{aliapoulios2021early}. In addition, the study reports that QAnon is the 8th most popular hashtag on Parler.

\subsubsection{Cross-platform studies}
Cross-platform studies allow for a fair comparison and deepen our understanding of the roles different platforms play in the ecosystem of information. The latter is particularly important when it relates to phenomena damaging to society (e.g. conspiracy theories) or to individuals (e.g. anti-social language). However, not many studies have been published in the context of conspiracy-related content, anti-social language or hate speech. Notable exceptions include an analysis and comparison of language related to QAnon, across sites collecting posts claimed to be written by ``Q'', Twitter, 4chan, 8chan, Reddit, and Voat \cite{aliapoulios2021gospel}. The study suggested that the QAnon community can find a home for their content on mainstream platforms, and that bans on one platform ``do little to slow growth on others''. The same study reports that posts written by ``Q'' are less toxic than posts by QAnon communities on sites such as Voat and 4chan, although the study does not capture the same measure for Twitter.

Several comparative studies have been published on social media engagement with the Covid-19 pandemic, some focusing on conspiratorial content. A study of Covid-19-related conspiracy theories on 8kun and Gab used 8kun's most prominent QAnon related board ``/qresearch'' to collect pandemic-related conspiracy theories \cite{zeng2021conceptualizing}. The study found that 24\% of the Covid-related posts on Gab contain conspiracy theories, and that 57\% of randomly selected user profiles contain conspiracy theory content, such as QAnon, although that the prevalence of such content is higher on 8kun. Another study has focused on the effect of moderation, contrasting Facebook, Twitter, Reddit, and 4chan \cite{papakyriakopoulos2020spread}, finding that ``content moderation on Twitter was less effective than on the other platforms'', potentially attributed to the fact that content on Twitter spreads within the first hours of it being posted. Finally, a study on the emergence of sinophobic behaviour on web communities in the face of the pandemic on Twitter and 4chan has focused on content analysis, including racist slurs, as a type of hate speech \cite{tahmasbi2021go}.

Other existing scholarship involving multiple platforms and hate speech includes a comparative study of hate speech on Twitter and Reddit around attacks involving Arabs and Muslims as perpetrators or victims \cite{olteanu2018effect}. The authors observed that ``extremist violence tends to lead to an increase in online hate speech'', with the biggest increase seen in messages advocating for the violence. A comparison of Gab and 4chan's Politically Incorrect board (/pol/) in the context of antisemitism, found evidence of increasing antisemitism around political events such as the US presidential election of 2016 \cite{zannettou2020quantitative}. Finally, Gab was reported to have a ``high percentage of news shared to be from untrustworthy sources (48.7\%), compared to 8.7\% for Twitter'' \cite{wang2021multi}.

\section{Data}
To compare the discourse related to the \#QAnon, we have collected the data from all three platforms, including all posts which have been posted with the \#QAnon hashtag (not case sensitive) in the period between December 7, 2020 (15:44:33) and January 10, 2021 (20:22:15) CET.

We obtained Parler data via Parlance API \cite{parlance}. While unofficial, we confirmed that the API has allowed us to access the same data that would have been accessible if we had used Parler's search feature to search for the same hashtag. The data collection started on January 6 and ended when Parler went offline, at 08:59 CET on January 12. We collected data for the same hashtag, for the same time period from Gab using garc library \cite{garc}, and from Twitter using Twitter's Academic full archive search API, via the academictwitteR library \cite{twitterAcademic2021, academictwitteR}. The data for all three platforms included the text of the post, as well as metadata (such as information on author, time, date).

\subsubsection{Sampling strategy}
\#QAnon is not the only hashtag used by QAnon supporters. A notable alternative, \#wwg1wga (``where we go one, we go all''), was reportedly more popular than \#QAnon on Parler, although in a sample that is not statistically representative \cite{aliapoulios2021early}. On the other hand, our experiments showed that \#QAnon is more popular than \#wwg1wga on Twitter. When a movement uses multiple hashtags, there is not a perfect choice of a hashtag to analyse. We choose \#QAnon as we believe that a person who is not well-versed or already a believer of the conspiracy theory would not know what \#wwg1wga is, and \#QAnon is a natural and easy first choice to search for. In addition, we highlight that our study aims to establish how platforms compare to each other, rather than how \#QAnon-related posts compare to users’ overall activity on each platform. We do not draw conclusions about how \#QAnon compares to non-\#QAnon content, or other hashtags. 

\subsubsection{On comparability of platforms}
For all three platforms, we analyse re-posts made with additional comments as they provide additional information, while excluding verbatim re-posts from the analysis. We make this choice as the data retrieval methods for Parler and Gab do not allow us to know what posts have been re-posted without additional comment within the time period of interest. Parler and Gab allow comments on posts, while Twitter does not. We exclude comments from the analysis to ensure comparability, although we note that language in comments may differ from the language in posts, which warrants a separate study.


\begin{table}[t]
    \centering
    \resizebox{\columnwidth}{!}{
    \begin{tabular}{lllll}
        \toprule
        Platform  & Collected & Analysed & Users & Posts per user ($\mu$)\\
        \midrule
        Twitter &  15 706 & 12 325 & 5861 & 2.18 \\
        Parler & 81 456 & 78 892 & 4648 & 17.52 \\
        Gab & 6 997 & 6 708 & 501 & 13.97 \\
        \bottomrule
    \end{tabular}}
    \caption{Number of collected vs analysed posts, number of users, and average number of posts per user. Posts which were collected, but not analysed, do not contain text in English.}
    \label{table:samplesizetable}
\end{table}

Hashtags are a feature on all three platforms, and their intended use is the same - to allow users to add keyword-like metadata to their posts, signifying a topic. All three platforms enable their users to add any number of hashtags to their posts - but the hashtags count within the character limit. Character limits vary across platforms - the maximum length of posts on Twitter, Parler and Gab being 280, 1000, and 3000 characters respectively. As such, hashtagging behaviour is different across the platforms, and the ratio of hashtags to text across the three data sets varies. While this has implications for our results, cross-platform research that is necessary to understand social media usage in perspective inevitably has to deal with the differences in platform functionalities and/or usage and account for them, as we do and describe below. Further, as the intended function of hashtags is the same across the three platforms, the comparison remains relevant despite these differences.

\subsection{Data-related limitations}
\subsubsection{Language}
Twitter API returns the information on the language of the post. For Parler and Gab, we have used Perspective API's automatic language detection to determine the language \cite{perspectiveFeatures}. In our data, 96.9\% of Parler, and 95.9\% of Gab are posts in English, and while Twitter contains a larger proportion of non-English posts (19\% in languages other than English), we limit our analysis to posts written in English, to ensure the comparability across the three platforms. Table \ref{table:samplesizetable} shows the number of posts that have been collected, and analysed, across the three platforms.

\subsubsection{Moderation}
Twitter has been moderating content related to QAnon since at least late July 2020 \cite{twitter2020}. The implication is that there likely is a body of Tweets that Twitter has removed and are not available for analysis. It is not unreasonable to assume that these Tweets might contain a higher amount of anti-social language, although that can't be established without an analysis made impossible due to the lack of data. Due to this, we consider anti-social language results on Twitter a lower bound for the platform. We collected Parler data over the last few days of Parler being online in January, and Gab data on February 28. While Gab reportedly has little to no moderation \cite{lima2018inside}, we can not guarantee that our data set captures everything that had been posted. Still, given the largely unmoderated nature of both Gab and Parler, we argue that it is reasonable to assume that only a small number of posts would have been removed, if any, and that our Parler and Gab data truthfully represent the conversation on those platforms, as it happened.

\subsubsection{Time period}
Our study focuses on a very specific time period. While this is partially driven by the fact that Parler going offline interrupted our data collection, it also allowed us to achieve higher internal validity of the data. Additionally, the month of analysis includes events that could increase interest in QAnon, as well as lead to increased activity on social media platforms. Conversely, this means that the volume of data collected is limited, which could affect analysis of language undertaken as part of RQ3 and RQ4. In future work, it could be relevant to examine the development of cross-platform QAnon-related discourse over time and, for instance, compare the observations during the period after the US presidential election, shortly before it and in more routine periods.


\section{RQ1: Activity, users, and posting style}

To understand how the volume of \#QAnon activity, number of users participating in the discussion, and posting styles vary across the three platforms, we utilise metadata extracted from posts, which allow us to summarise temporal posting activity, number of users posting, and similar descriptive statistics. In addition to analysing the metadata, we extracted various features from the text, such as hashtags, and text length descriptive statistics, to capture differences in posting styles.

Table \ref{table:samplesizetable} shows the volume of posts with \#QAnon for the observation period on each platform. On Parler, the number of posts was much higher than on Gab and Twitter. Though at first glance that indicates a much higher prevalence of \#QAnon-related discussions on Parler as compared to Gab and Twitter, the observation might be partially explained by the differences in the way hashtags are used across platforms.

\begin{table}[t]
    \centering
    \resizebox{\columnwidth}{!}{
    \begin{tabular}{lccc|cc|cc}
        \cmidrule{2-8}
        \multirow{2}{*}{} & \multicolumn{3}{c}{Characters} & \multicolumn{2}{|c|}{Words} & \multicolumn{2}{c}{Hashtags} \\
        \cmidrule(lr){2-4} \cmidrule(lr){5-6} \cmidrule(lr){7-8}
                &  max   &  $\mu$  & $\sigma$& $\mu$  &$\sigma$& $\mu$  & $\sigma$ \\
        \midrule
        Twitter &  280   &  154.8  &  82.3   &  23.3  &  13.3  &  4.6   & 4.7 \\
        Parler  &  1000  &  561.6  &  310.9  &  56.9  &  31.7  &  39.3  & 25.3 \\
        Gab     &  3000  &  512.6  &  456.3  &   61   &  59.2  &  26.3  & 31.7 \\
        \bottomrule
    \end{tabular}}
    \caption{Means ($\mu$) and standard deviations ($\sigma$) for number of characters, words and hashtags in posts}
    \label{table:poststats}
\end{table}

\begin{figure}[t]
    \centering
    \includegraphics[width=0.9\columnwidth]{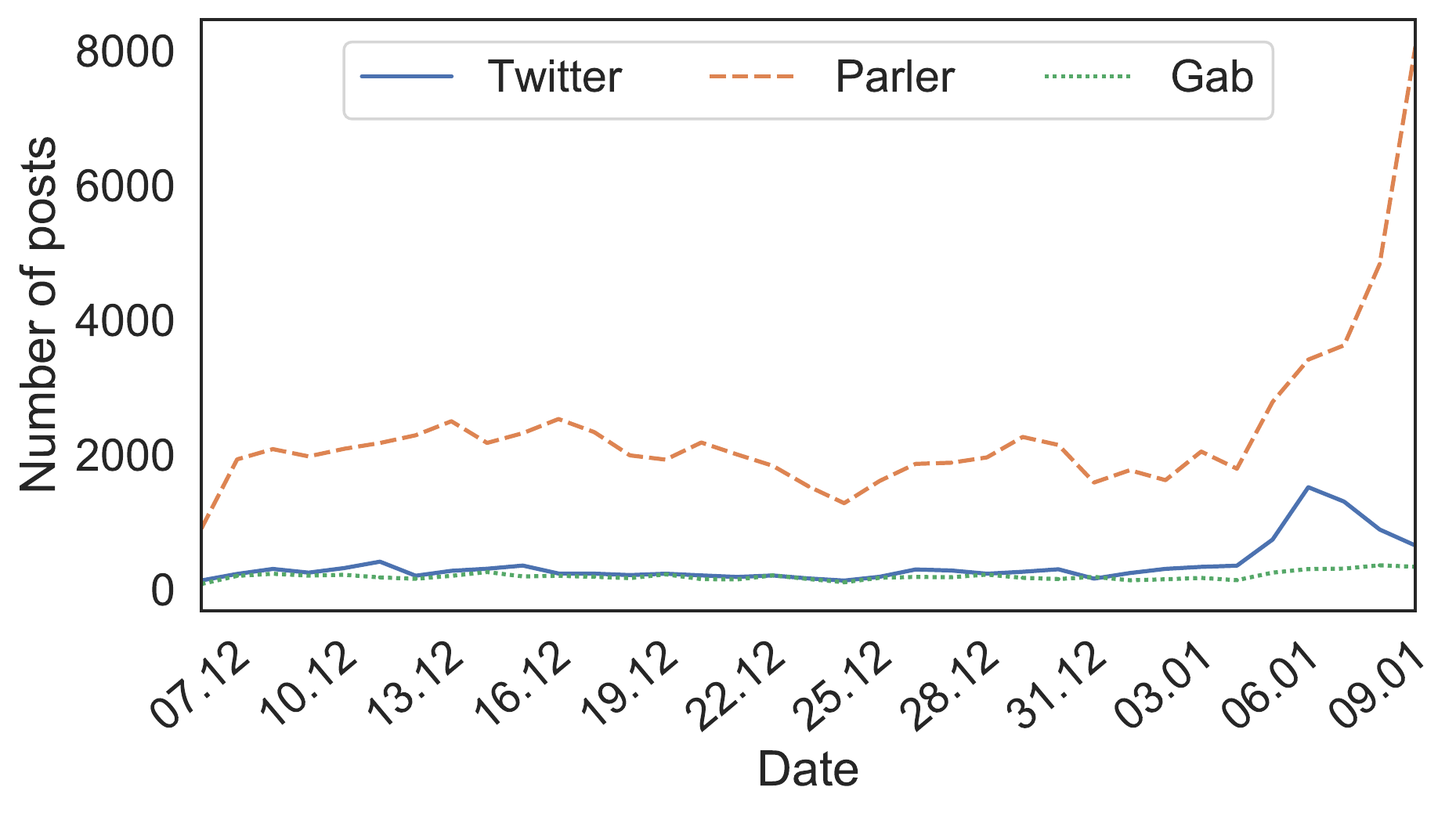}
    \caption{Daily frequency of \#QAnon posts}
    \label{fig:postfrequency}
\end{figure}


The discrepancies in the number of users who posted about \#QAnon between platforms were smaller than the differences in the sheer volume of posts with this hashtag. Though Parler contained substantially more posts with \#QAnon than Twitter, the number of unique users posting with the hashtag was higher on the latter (Table \ref{table:samplesizetable}). This difference might mean that Parler and Gab users include this hashtag in posts not explicitly about QAnon, or are more invested in QAnon-related discussion than Twitter users once involved.

Average account ages on the three platforms differ as well. The average account age at the time of posting on Twitter was 2781 days, on Parler 131 days, and on Gab 675 days. To examine whether this discrepancy is attributed to the difference in the age of the platform alone (Twitter was founded in 2006, Gab in 2016, and Parler in 2018) or to external events, we further scrutinised the dates when users joined each platform. We found that 19.8\% of \#QAnon posts on Parler and 18.8\% on Gab were created by users who have been on the platform for less than 1 month. On Twitter, the share of such users is only 1.7\%. This implies that the average account age is not attributed only to the age of the platform but also external events such as the election and events that followed: of the users posting with the \#QAnon, 1.9\% have joined Twitter after the Election (November 3, 2020), compared to 56.1\% on Parler, and 40.1\% on Gab. 

However, the apparent growth in the number of users who joined Gab and Parler during the one-month observation period was not accompanied by the bursts of \#QAnon posting activity. The number of posts with the hashtag was relatively stable across the three platforms up until early January, as depicted in Figure \ref{fig:postfrequency}. Then both Twitter and Parler but not Gab saw a rise in popularity of the hashtag around January 6, with the conversation on Parler further intensifying in volume in the run-up to the platform going offline.

In Table \ref{table:poststats} we report the means and standard deviations for character counts, word counts, and the number of hashtags. Unsurprisingly, posts on Gab and Parler are longer on average as the two platforms have higher character limits than Twitter. The higher limit leaves the users of the two ``alt-tech'' platforms more space for hashtags, with the average number of hashtags being much higher on Gab and Parler than on Twitter, and 75.5\% of Parler posts having more hashtags than other words, compared to 51.7\% for Gab, and 17.5\% for Twitter. The extra character limit is employed for particularly active hashtagging by Parler users. The mean words to hashtag ratio on Parler is 1.44, compared to 6.94 on Gab and 10.15 on Twitter. This points to a very different way in which hashtags are used across the platforms. In fact, upon a qualitative inspection of a sample from all three, we noticed that many Parler posts used ``hashtag walls'' - blocks of many continuous hashtags, not necessarily related to the post itself. The observed propensity of Parler users to include a high number of hashtags, not always related to the immediate content of the post, might partially explain why the volume of posts on Parler with \#QAnon is much higher than on Twitter and Gab.

\section{RQ2: Anti-social language}
To compare the three platforms with respect to the prevalence of anti-social language, we capture measures of anti-social language and hate speech, using similar methodology as many other studies - Perspective API \cite{pavlopoulos2020toxicity, elsherief2018hate, aliapoulios2021gospel, zannettou2020measuring} and Hatebase lexicon of hate words \cite{silva2016analyzing, mcilroy2019welcome, zannettou2018gab}.

We used Perspective API to extract features that act as proxies of anti-social language \cite{perspectiveFeatures}. While it has exhibited bias in certain contexts (e.g. classifying language predominantly used by African-Americans as more toxic than the language used by white people \cite{sap2019risk}), it was also reported to outperform other available tools on similar texts \cite{zannettou2020measuring}. For each feature of interest, when queried against text, Perspective provides a score between 0 and 1, denoting the probability that the post is as the feature describes. As it is a probability, a threshold that is set to determine if a post is, for example, severely toxic, is both arbitrary and set according to the application of interest. In our comparison, rather than choosing an arbitrary threshold, we look at the distribution of scores to capture the differences. We perform statistical testing using a two-sample Kolmogorov-Smirnov test to test for the difference in score distributions. All results we report below are statistically significant at p \textless 0.001. The features we have selected to analyse are presented in Table \ref{table:perspectivetable}.


\begin{table}[t]
    \centering
    \resizebox{.9\columnwidth}{!}{
    \begin{tabular}{p{0.21\linewidth} p{0.75\linewidth}}
        \toprule
        Feature  & Description \\
        \midrule
        Severe \newline toxicity & A very hateful, aggressive, disrespectful comment or otherwise very likely to make a user leave a discussion \\ \midrule
        Threat & Describes an intention to inflict pain, injury, or violence against an individual or group \\ \midrule
        Identity \newline attack & Negative or hateful comments targeting someone because of their identity \\ \midrule
        Insult & Insulting, inflammatory, or negative comment towards a person or a group of people \\
        \bottomrule
    \end{tabular}}
    \caption{Perspective feature descriptions \cite{perspectiveFeatures}}
    \label{table:perspectivetable}
\end{table}

\begin{figure}[t]
    \centering
    \includegraphics[width=0.9\columnwidth]{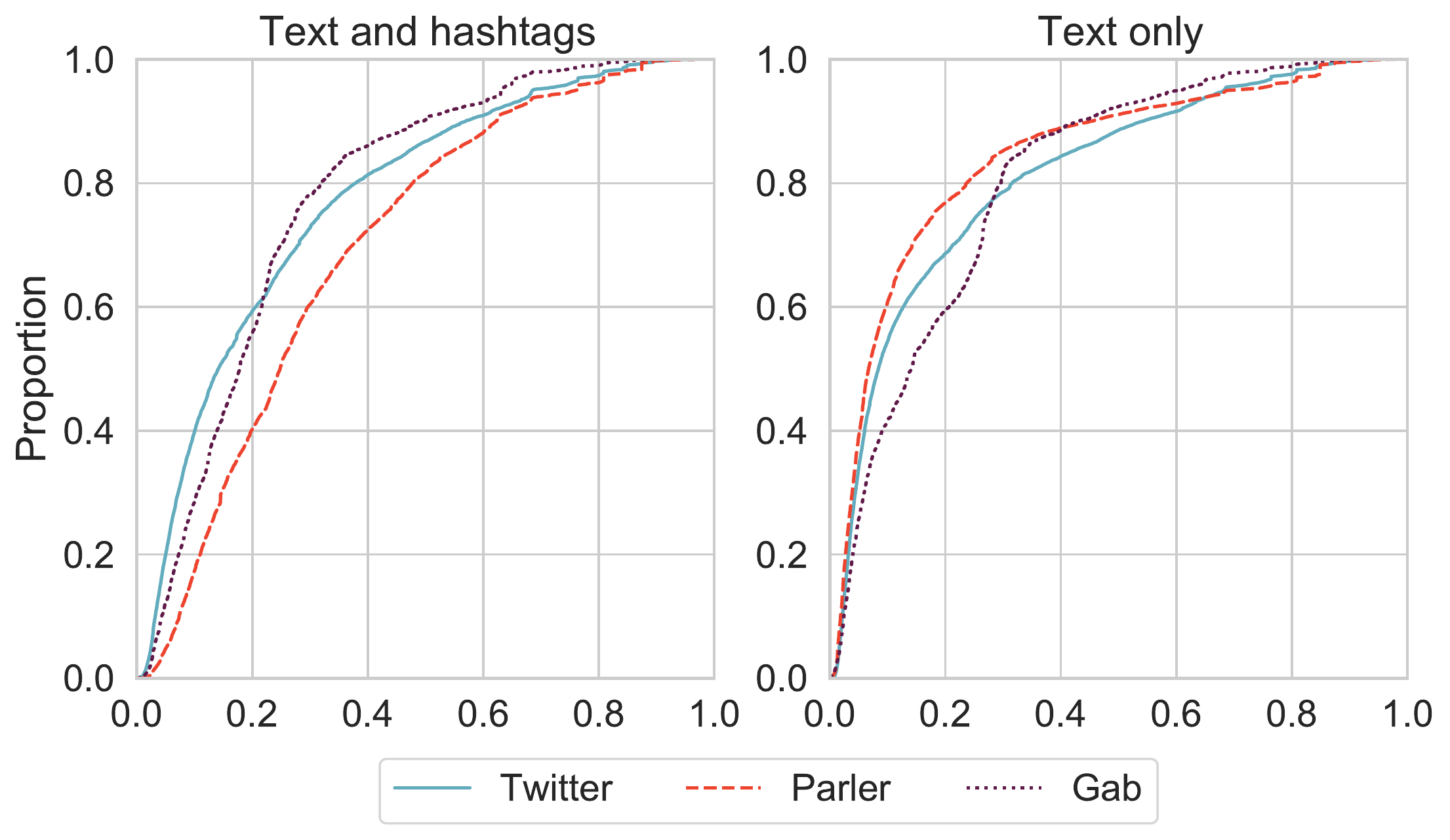}
    \caption{CDFs of severe toxicity for text only, and for text and hashtags combined.}
    \label{fig:sevtox_oldvnew}
\end{figure}

To examine the effect of hashtags on anti-social language measures, due to the difference in hashtagging behaviour across the platforms discussed in the previous section, we query Perspective three times for each post. Firstly, we query it on the post cleaned of URLs and mentions, secondly having additionally removed all hashtags, and finally on hashtags alone. The cross-platform differences in hashtagging have important implications for the results in relation to the prevalence of anti-social language. In Figure \ref{fig:sevtox_oldvnew}, we show cumulative distributions of severe toxicity scores across the three platforms - the plot on the left showing scores of whole posts, and the plot on the right showing severe toxicity of text part of the post only, with hashtags removed. The figure shows that without the hashtags, Parler shows lower or similar toxicity to Twitter and Gab. However, once hashtags are added, Parler consistently scores worse.

\begin{figure}[t]
    \centering
    \includegraphics[width=0.9\columnwidth]{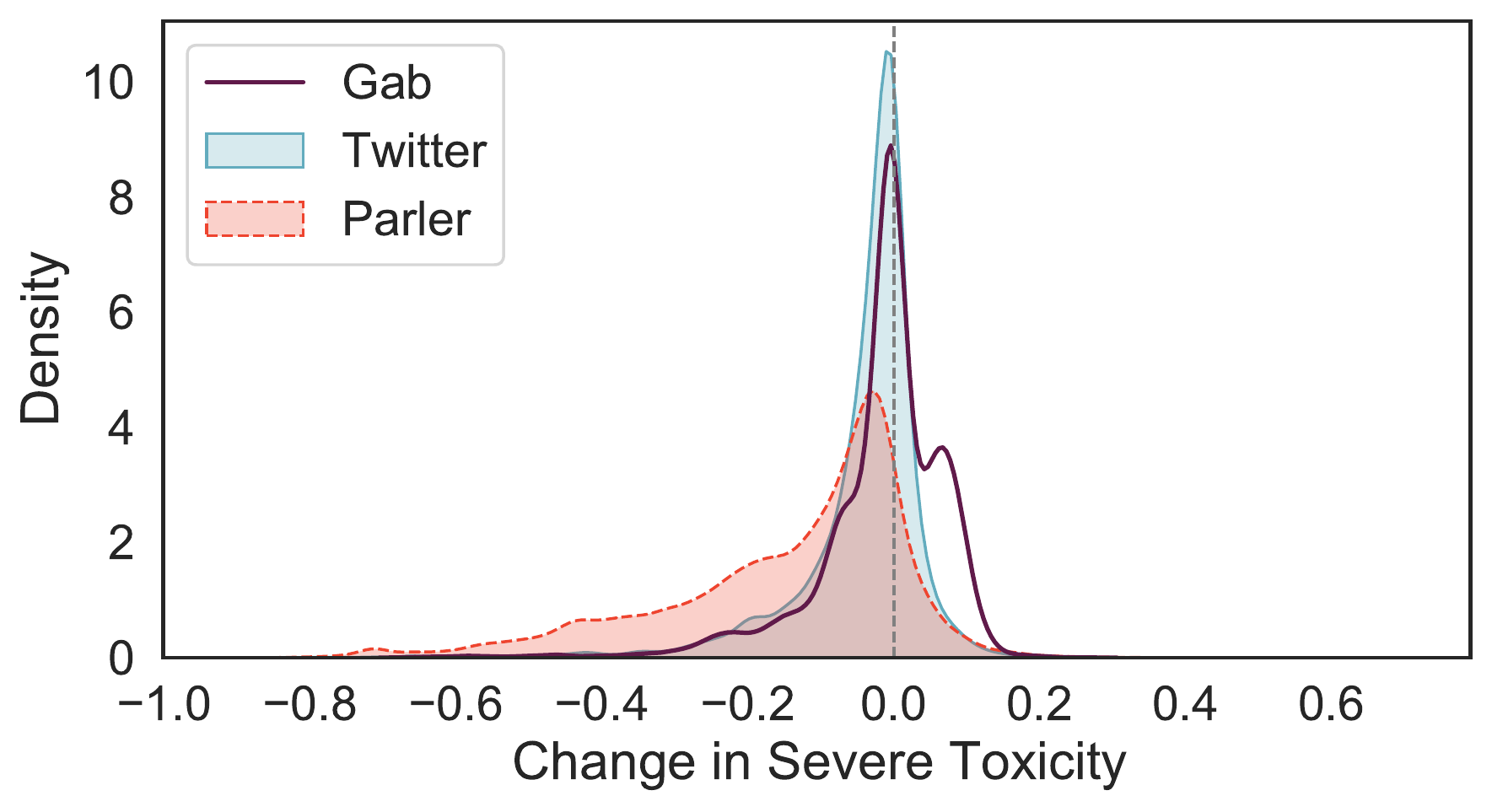}
    \caption{Impact of adding hashtags on severe toxicity. Negative values indicate that a post is more toxic with hashtags.}
    \label{fig:hashNoHashSevToxDiff}
\end{figure}

\begin{figure}[t]
    \centering
    \includegraphics[width=\columnwidth]{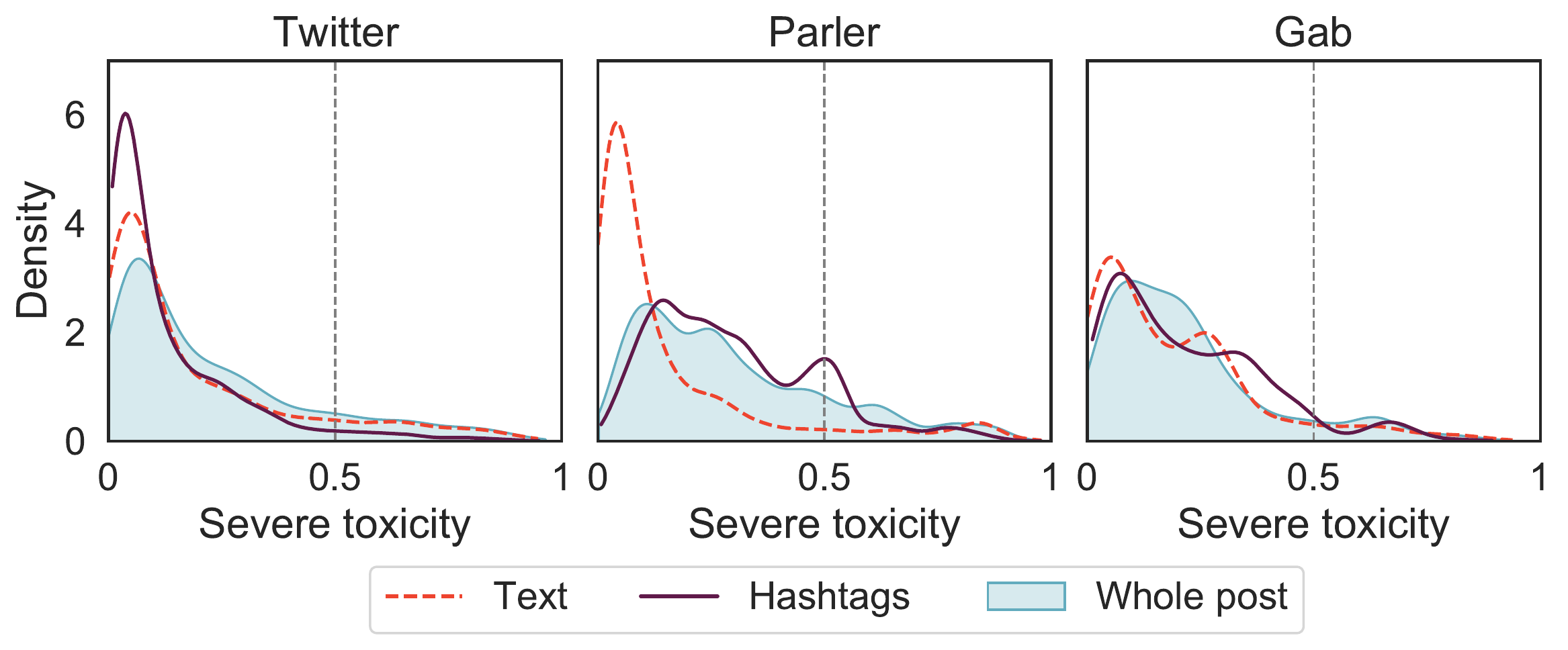}
    \caption{A kernel density estimate of severe toxicity of different post components - hashtags, text, and hashtags and text combined.}
    \label{fig:sevtox_textvtags}
\end{figure}

We further explored the relationship between the presence of hashtags and the posts' classifications. In Figure \ref{fig:hashNoHashSevToxDiff}, we show the distribution of differences in Severe Toxicity scores between full posts and posts with tags removed, by platform. Values smaller than 0 indicate that a post is assigned a higher severe toxicity score with hashtags. As a much longer left tail for Parler indicates, Parler posts with hashtags included are assigned much higher severe toxicity scores compared to the same posts without hashtags. Severe toxicity scores for different post components across all three platforms are in Figure \ref{fig:sevtox_textvtags}. Parler's hashtagless texts, compared with hashtags only or combined (texts and hashtags), are overwhelmingly not severely toxic, and compare to scores of tweets.

\begin{figure}[t]
    \centering
    \includegraphics[width=\columnwidth]{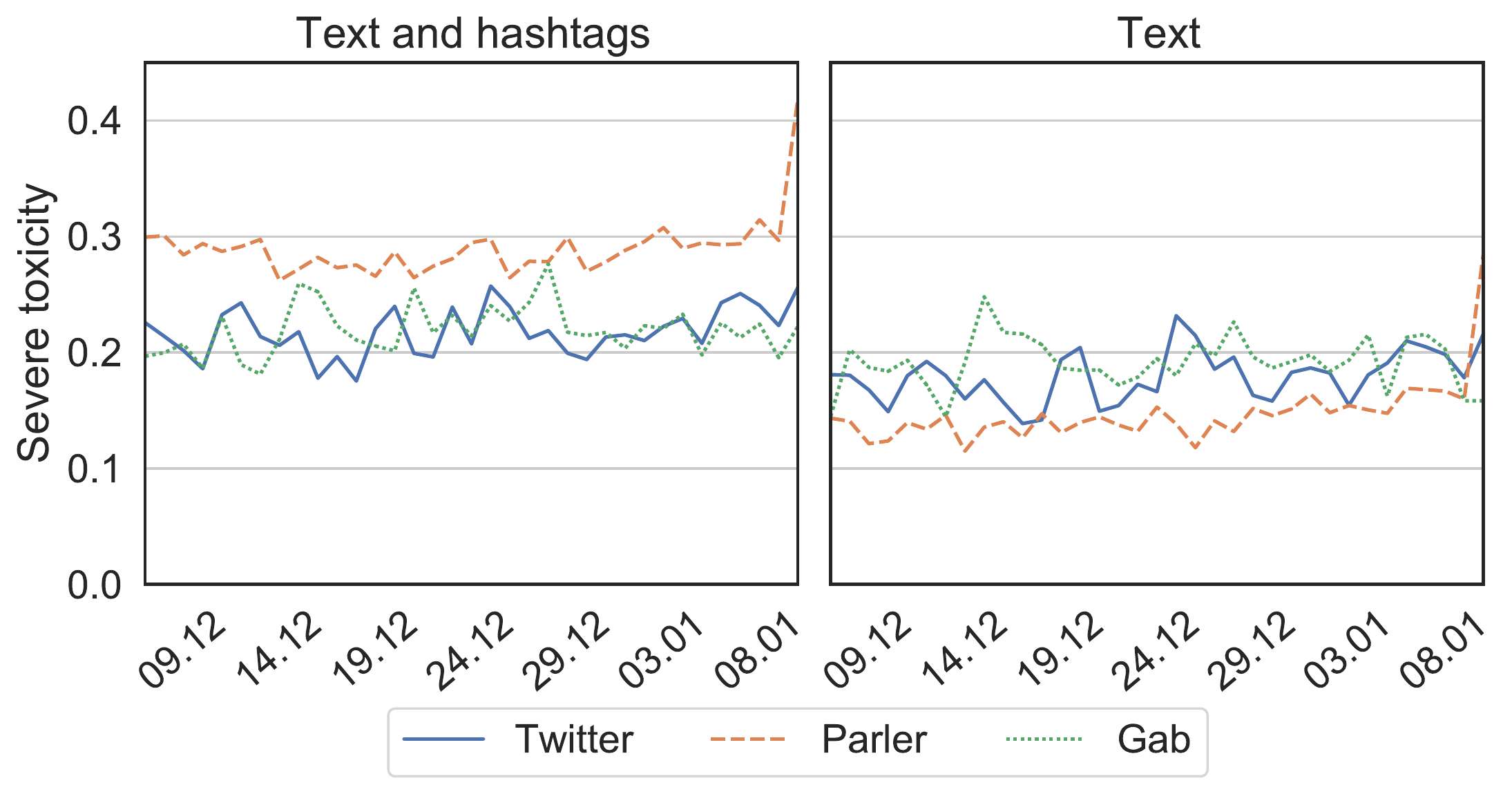}
    \caption{Daily mean severe toxicity of posts, with hashtags, and without}
    \label{fig:sevtoxTemporal}
\end{figure}

Mean scores for severe toxicity across time on all three platforms are presented in Figure \ref{fig:sevtoxTemporal}. The addition of hashtags changes the overall picture in this case as well. For full posts (with hashtags), Parler consistently has a higher mean severe toxicity score assigned than Twitter and Gab, which have similar scores to each other. On the other hand, with the hashtags removed, Parler consistently has lower mean severe toxicity scores than both Twitter and Gab, except for a large uptick in toxicity seen after January 6.

\begin{figure}[t]
    \centering
    \includegraphics[width=\columnwidth]{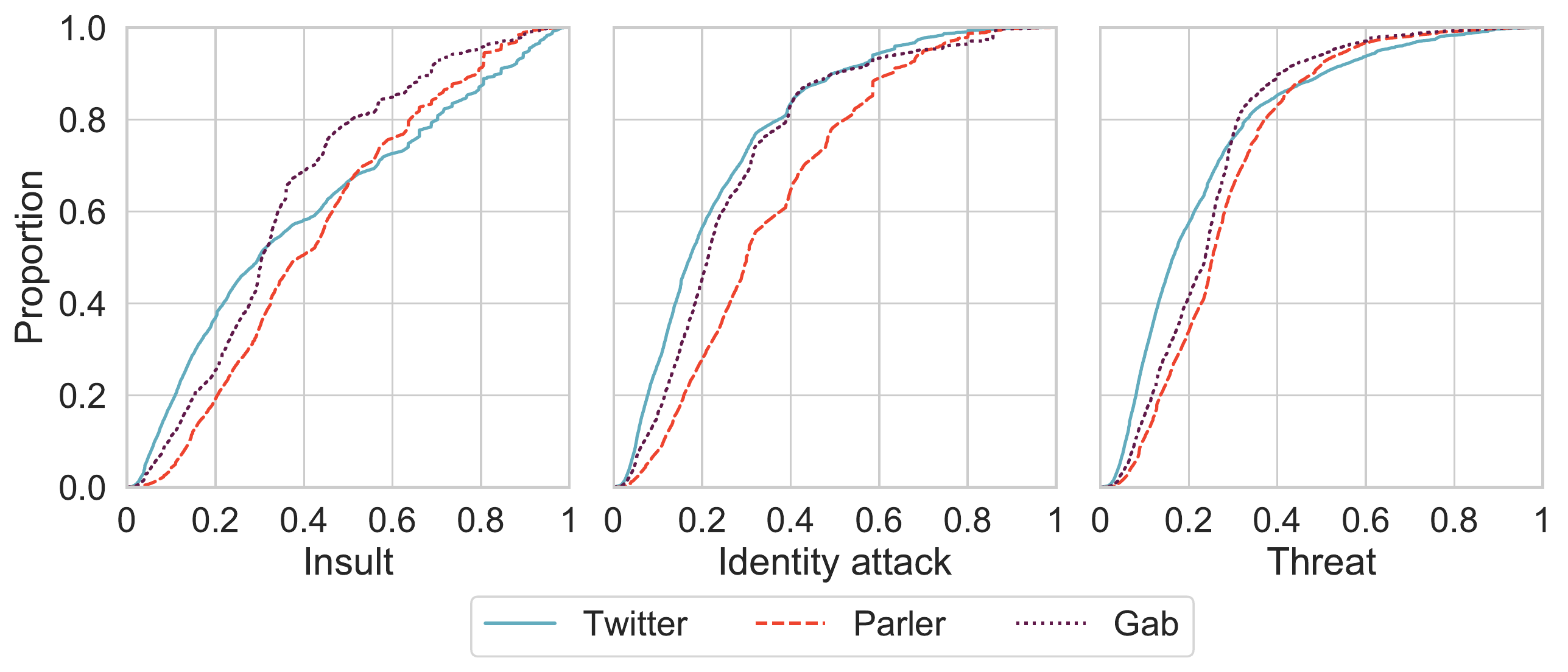}
    \caption{CDFs for Insult, Identity attack, and Threat models}
    \label{fig:perspectiveCDFs}
\end{figure}

Cumulative distribution functions (CDFs) for Insult, Identity attack, and Threat features are presented in Figure  \ref{fig:perspectiveCDFs}. 
The model identifying threatening language shows that posts on all three platforms do not have a high probability of being threatening, with platforms having a similar proportion of posts more likely than not to be threatening (with a probability over 0.5). The identity attack model shows Parler posts scoring the worse, with Twitter and Gab being very similar. Finally, the platforms only show minor differences when it comes to insults, although a lower proportion of Gab posts are likely to be insulting than posts on Twitter and Parler. 

While Perspective allows us to capture measures such as toxicity, and it takes into account words that are considered hate speech, we separately measure hate speech specifically, as the most extreme type of anti-social language. We perform keyword-based classification by using Hatebase lexicon of hateful terms \cite{hatebase2021}. Hatebase collects words and phrases considered hateful across multiple categories, such as ethnicity, nationality, religion, gender, class, etc. We matched cleaned text (without links, mentions, or hashtags) from all three platforms with words and phrases in the lexicon.

Upon inspection, a few ambiguous words have resulted in a high number of false positives (e.g. ``Apple'' is in Hatebase lexicon as it can be used to signify ``An American Indian who is 'red on the outside, white on the inside.''', but all occurrences of the word in our corpus referred to the company Apple Inc., or the fruit apple). For this reason, we have omitted a number of words\footnote{ABC, Afro-Saxon, Anglo, Ann, apple, banana, Becky, bird, boos, bubble, bucks, Charlie, chief, coconuts, egg, egg-plant, frog, girl, guinea, lefties, mock, pancakes, pepper, Pepsi, property, queen, skinny, snowflake, sole, spikes, Tommy, Yankee, yellow} from the analysis. We chose words to remove by manually going through each occurrence of a hate word, and removing it from a dictionary if all occurrences of it were used in a non-hateful context. While Gab scores similarly, or even better, on different anti-social language measures than Twitter and Parler, it takes the leading position in terms of the share of posts with hate words (see Table \ref{table:hatebaseResults} for the exact percentages and corresponding numbers of posts).

\begin{table}[t]
    \centering
    \resizebox{.55\columnwidth}{!}{
    \begin{tabular}{lll}
        \toprule
        \multirow{2}{*}{Platform} & \multicolumn{2}{c}{Posts with hate words} \\
        \cmidrule{2-3}
         & Count & Percentage \\
        \midrule
        Twitter & 345 & 2.8 \% \\
        Parler & 2098 & 2.66 \% \\
        Gab & 320 & 4.77 \% \\
        \bottomrule
    \end{tabular}}
    \caption{Hatebase results}
    \label{table:hatebaseResults}
\end{table}

We conclude by noting that most indicators show that the language of most posts across all three platforms should not be considered anti-social. An exception is a higher probability of posts being insulting on Twitter and Parler. Language on Parler appears marginally worse in terms of severe toxicity and identity attacks than the other two platforms, although we have demonstrated that this result is greatly affected by the prolific hashtagging of Parler users. While most Gab posts have a lower probability of being anti-social than Parler, and even Twitter for some indicators, they also have the highest percentage of hate speech occurrence.

\section{RQ3: Themes and political figures}
To compare similarities and differences in mentioned themes (e.g. individuals, locations, phrases), we use the NLP library Stanza's named entity recognition feature \cite{qi2020stanza} on clean text (after removing links, mentions, and hashtags). We discarded some categories of entities, such as numbers and percentages, as they were of no interest for the analysis. The resulting entity list required cleaning, as terms signifying the same entity are not always automatically recognised as the same (e.g., ``January 6'', ``Jan 6'', ``6. January''). We calculate the frequency of occurrence of entities across platforms, and compare their relative popularity.

Following the removal of irrelevant entities, we have obtained 12759 named entities from Twitter, 90769 from Parler, and 15937 from Gab. 50.8\% of Twitter posts, 42.7\% of Parler posts, and 54.7\% of Gab posts contain at least one entity. The number of unique entities extracted from Twitter is 3937, from Parler is 17439, and from Gab is 5066, suggesting that the conversation on Twitter is more focused around a few subjects, while on Parler and Gab there is a higher diversity of entities discussed in connection to \#QAnon.


\begin{table}[t]
    \centering
    \resizebox{\columnwidth}{!}{
    \begin{tabular}{llll}
        \toprule
        Rank & Twitter & Parler & Gab \\
        \midrule
        1 &  Donald Trump & Donald Trump  & United States  \\
        2 &  United States & United States  & Donald Trump  \\
        3 &  QAnon & Joe Biden  &  Joe Biden \\
        4 & Republican  & Democrat  & Democrat  \\
        5 & Georgia  & Republican  & today  \\
        6 & GOP  & China  & China  \\
        7 & today  &  American &  Washington D.C. \\
        8 & Twitter  & Washington D.C.  &  patriots \\
        9 & american & Georgia  &  Twitter  \\
        10 & Joe Biden  & 2020  & FBI  \\
        11 & americans  & Twitter  & 2020  \\
        12 & Capitol  & today  &  Gab \\
        13 & Washington D.C.  & patriots  & congress  \\
        14 & senate  & January 6  &  Mike Pence \\
        15 & antifa  & chinese  & american  \\
        16 & Democrat  &  Mike Pence & Lin Wood  \\
        17 & Jake Angeli  & Nancy Pelosi  &  Georgia \\
        18 & Mike Flynn  & americans  &  CCP \\
        19 & yesterday  & Parler  &  tomorrow \\
        20 & Sidney Powell  &  GOP & QAnon \\
        \bottomrule
    \end{tabular}}
    \caption{Top 20 terms on the three platforms}
    \label{table:top20NER}
\end{table}

The top 20 entities from all three platforms are shown in Table \ref{table:top20NER}. Many entities are popular across all platforms (Donald Trump, United States, Georgia, Twitter, Joe Biden, American(s), Democrat). Even the terms which appear in the top 20 on only two of the three platforms (e.g. ``Republican'' and ``GOP'' on Twitter and Parler) are still just outside of the top 20 on the third (25th, and 33rd most popular on Gab, respectively). This suggests that the discourse in relation to \#QAnon on all three platforms was centred largely around similar subjects.

To infer the differences between popular entities, we first removed entities used by less than 10 users (on either platform). This is to filter out unusual and erroneous entities which are used by one user repeatedly (e.g. ``q and the plan to save the world''). We present the top 30 entities that are relatively more popular on one platform (in the top 50 of the most popular entities), as compared to the other two, in Table \ref{table:newDifference}. If $r(x)_t, r(x)_p, r(x)_g$ represent the ranking of popularity of term $(x)$ on Twitter, Parler and Gab, the maximum difference is calculated as $$\max(|r(x)_t-r(x)_p|, |r(x)_t-r(x)_g|, |r(x)_p-r(x)_g|)$$ The results are indicative of potential differences in the focus of the discussions on the three platforms. For instance, mentions of the Capitol, Jake Angeli (``QAnon Shaman''), Antifa, Nazis as well as Kelly Loeffler and Marjorie Taylor Greene were more prevalent on Twitter than on Parler or Gab, while the latter two platforms saw higher popularity of Pennsylvania (probably connected to the vote count in the state) and Dominion voting machines (according to conspiracy theories, a company which aided the ``stealing'' of the election), as well as Deep state (Gab), MSM (``mainstream media'', Gab), 2nd Amendment (Parler) and defense (Gab).

\begin{table}[t]
    \centering
    \resizebox{\columnwidth}{!}{
    \begin{tabular}{p{0.21\linewidth} lll | p{0.25\linewidth} lll}
        \toprule
        \multirow{2}{*}{Entity} & \multicolumn{3}{c|}{Rank} & \multirow{2}{*}{Entity} & \multicolumn{3}{c}{Rank}\\
        \cmidrule{2-4} \cmidrule{6-8}
        & Twi & Par  & Gab  & & Twi & Par  & Gab \\
        \midrule
        K. Loeffler  & \textbf{49} & 84 & 634          & Texas & 134 & \textbf{23} & \textbf{21} \\
        M.T. Greene  & \textbf{44} & 563 & 590         & Michigan     & 172 & \textbf{36} & 59 \\
        J. Angeli    & \textbf{17} & 412 & 362         & Pennsylvania & 89 & \textbf{31} & \textbf{38}  \\
        A. Babbitt   & \textbf{35} & 174 & 109         & California & 101 & \textbf{44} & 79  \\
        A. Jones     & \textbf{37} & 208 & 132         & Russian & \textbf{31} & 81 & 88  \\ 
        J. Epstein   & \textbf{32} & 74 & 140          & CCP & 152 & \textbf{28} & \textbf{18} \\
        JFK          & \textbf{46} & 150 & 155         & Nazi & \textbf{41} & 130 & 118 \\ 
        Dave         & 271 & 506 & \textbf{45}         & 2. Amendment & - & \textbf{46} & 349 \\
        Brian        & 498 & 376 & \textbf{47}         & Elec. college & \textbf{43} & 129 & 65 \\
        Gab          & 551 & 54 & \textbf{12}          & DOJ  & 164 & 90 & \textbf{40} \\
        Amazon       & 146 & 79 & \textbf{50}          & Constitution & 80 & \textbf{45} & \textbf{25}  \\
        The republic & 404 & \textbf{32} & 240         & defense & - & 159 & \textbf{30} \\ 
        Deep State   & 170 & 195 & \textbf{23}         & Antifa & \textbf{15} & \textbf{30} & 72 \\
        Dominion     & 129 & \textbf{34} & \textbf{32} & Capitol & \textbf{12} & 66 & 60 \\
        MSM          & 82 & 58 & \textbf{29}           & Yesterday & \textbf{19} & \textbf{47} & 76  \\
        \bottomrule
    \end{tabular}}
    \caption{30 terms with the biggest difference in popularity}
    \label{table:newDifference}
\end{table}

To analyse if there is a difference in the use of anti-social language when discussing selected groups of political figures, we examined entities occurring at least 20 times in the whole corpus manually, and classified them into groups of interest. We form four groups, with political figures divided by gender and party\footnote{excluding Trump and Biden, who are analysed separately} (Table \ref{table:politicianSplit}). We consider either the party membership, or service in an administration, when dividing by party lines. As in RQ2, we ensure that we only report results stemming from different distributions (tested using a two-sample K-S test, at p\textless0.001, excluding posts that include both groups being compared to ensure the independence of samples).


\begin{table}[t]
    \centering
    \resizebox{\columnwidth}{!}{
    \begin{tabular}{p{0.1\linewidth} p{0.88\linewidth}}
        \toprule
        Group & Politicians \\
        \midrule
        Female (R) & M.T. Greene, K. Loeffler, J. Ellis, S. Powell, I. Trump, M. Trump \\ \midrule
        Female (D) & H. Clinton, N. Pelosi, K. Harris, M. Obama, A. Ocasio-Cortez, J. Biden, S. Abrams \\ \midrule
        Male (R) & M. Pompeo, T. Cruz, M. Pence, Bush, J. Hawley, M. Brooks, D. Perdue, R. Giuliani, M. McConnell, G. Sterling, M. Flynn, B. Barr, L. Graham, D. Scavino, R. Paul, J. Ratcliffe, K. McCarthy, C. Miller, R. Stone, B. Raffensperger, B. Kemp, M. Romney, S. Bannon, M. Gaetz, J. Jordan \\ \midrule
        Male (D) & B. Obama, J. Podesta, H. Biden, C. Schumer, A. Cuomo, E. Swalwell, G. Newsom, B. Clinton, A. Schiff, R. Warnock \\
        \bottomrule
    \end{tabular}}
    \caption{US Political figures mentioned at least 20 times}
    \label{table:politicianSplit}
\end{table}

Differences are observed with regard to the groups of political figures and two presidential candidates (Table \ref{table:termStats}). While Donald Trump and male Republican politicians were mentioned consistently more (in terms of the share of posts mentioning them) than their Democratic counterparts across all platforms, there was a divergence in posts about female politicians. There was a higher share of posts about female Republicans than Democrats on Twitter, while on Gab and Parler, the situation was reversed.

We observe significant differences in the mean Perspective scores of posts mentioning different groups of politicians (female vs male; Republican vs Democrat). On all three platforms, posts mentioning female (vs male) politicians, Democrats (vs Republicans) and Trump (vs Biden) scored higher on average for anti-social language features that exhibited significant cross-group differences (Table \ref{table:merged_ner_diff}).

\begin{table}[t]
    \centering
    \resizebox{\columnwidth}{!}{
    \begin{tabular}{p{0.35\linewidth}llllll}
        \toprule
        \multirow{2}{*}{Group} & \multicolumn{3}{c}{Percent of posts} & \multicolumn{3}{c}{Number of posts} \\
        \cmidrule(lr){2-4} \cmidrule(lr){5-7} 
                           &   Twi    &   Par    &   Gab    &   Twi   &   Par   &  Gab  \\
        \midrule
        Female Republicans &   1.18   &   0.62   &   0.66   &   145   &   487   &   44 \\
        Female Democrats   &   0.5    &   1.21   &   1.95   &   62    &   956   &   131 \\
        Male Republicans  &   2.08   &   2.69   &   3.91   &   256   &   2126  &   262 \\
        Male Democrats      &   0.54   &   1.3    &   1.43   &   66    &   1022  &   96 \\
        Donald Trump       &   9.71   &   9.6    &   8.59   &   1197  &   7572  &   576 \\
        Joe Biden          &   0.95   &   2.54   &   2.76   &   117   &   2000  &   185 \\
        \bottomrule
    \end{tabular}}
    \caption{Percentage, and the number, of posts containing mentions of politician groups (excluding Trump and Biden)}
    \label{table:termStats}
\end{table}

\begin{table}[t]
    \centering
    \resizebox{\columnwidth}{!}{
    \begin{tabular}{ll ll ll ll}
        \cmidrule[\heavyrulewidth]{2-8}
        & \multirow{2}{*}{Feature} & \multicolumn{2}{c}{Party} & \multicolumn{2}{c}{Gender} & \multicolumn{2}{c}{Candidate} \\
        \cmidrule(lr){3-4} \cmidrule(lr){5-6} \cmidrule(lr){7-8}
         &   &$\mu$(D)&$\mu$(R) &$\mu$(F)&$\mu$(M) &$\mu$(B)&$\mu$(T)\\  
        \midrule
          \multirow{3}{*}{Twi}&  Ide.Att. &  0.23  &  0.21   &     -    &      -   &  -     &    -       \\
           &  Insult   &    -    &   -      &     -    &     -    &  0.44 &      0.50 \\ \midrule
          \multirow{6}{*}{Par} &  Ide.Att. &  0.35  &  0.29   &    0.35 &   0.30  &  0.33 &      0.34 \\
           &  Insult   &  0.47  &  0.41   &    0.48 &   0.41  &  0.42 &      0.51 \\ 
           &  S. Toxic &  0.30  &  0.25   &    0.31 &   0.26  &  0.26 &      0.32\\ 
           &  Threat   &  0.26  &  0.24   &  -  &      -   &  0.24 &      0.26  \\ \midrule
           \multirow{5}{*}{Gab} &  Ide.Att.  &  0.29  &  0.20   &    0.26 &   0.23  &   -    &     -      \\ 
           &  Insult   &  0.43  &  0.33   &    0.45 &   0.34  &   -    &     -       \\ 
           &  S. Toxic &  0.21  &  0.16   &    0.20 &   0.17  &   -    &     -       \\ 
           &  Threat   &   -    &    -     &   -      &     -    &  0.22 &       0.26 \\
        \bottomrule
    \end{tabular}}
    \caption{Mean Perspective scores mentioning Democrats (D) or Republicans (R), female (F) or male (M) politicians, and Biden (B) or Trump (T).}
    \label{table:merged_ner_diff}
\end{table}

In conclusion, the entity-based analysis shows that while most prevalent named entities across the platforms are similar, there are differences between the mainstream platform and the alt-tech platforms. Twitter saw more mentions of high profile individuals considered QAnon supporters (Jake Angeli, Marjorie Taylor Greene), while more Parler and Gab posts used conspiratorial terms (Dominion, Deep state, Mainstream media). On all three platforms, posts mentioning female politicians, Democrats and Donald Trump score higher on all anti-social language features than posts mentioning male politicians, Republicans, and Joe Biden.

\section{RQ4: Narrative analysis}
\begin{figure}[t]
    \centering
    \includegraphics[width=0.8\columnwidth]{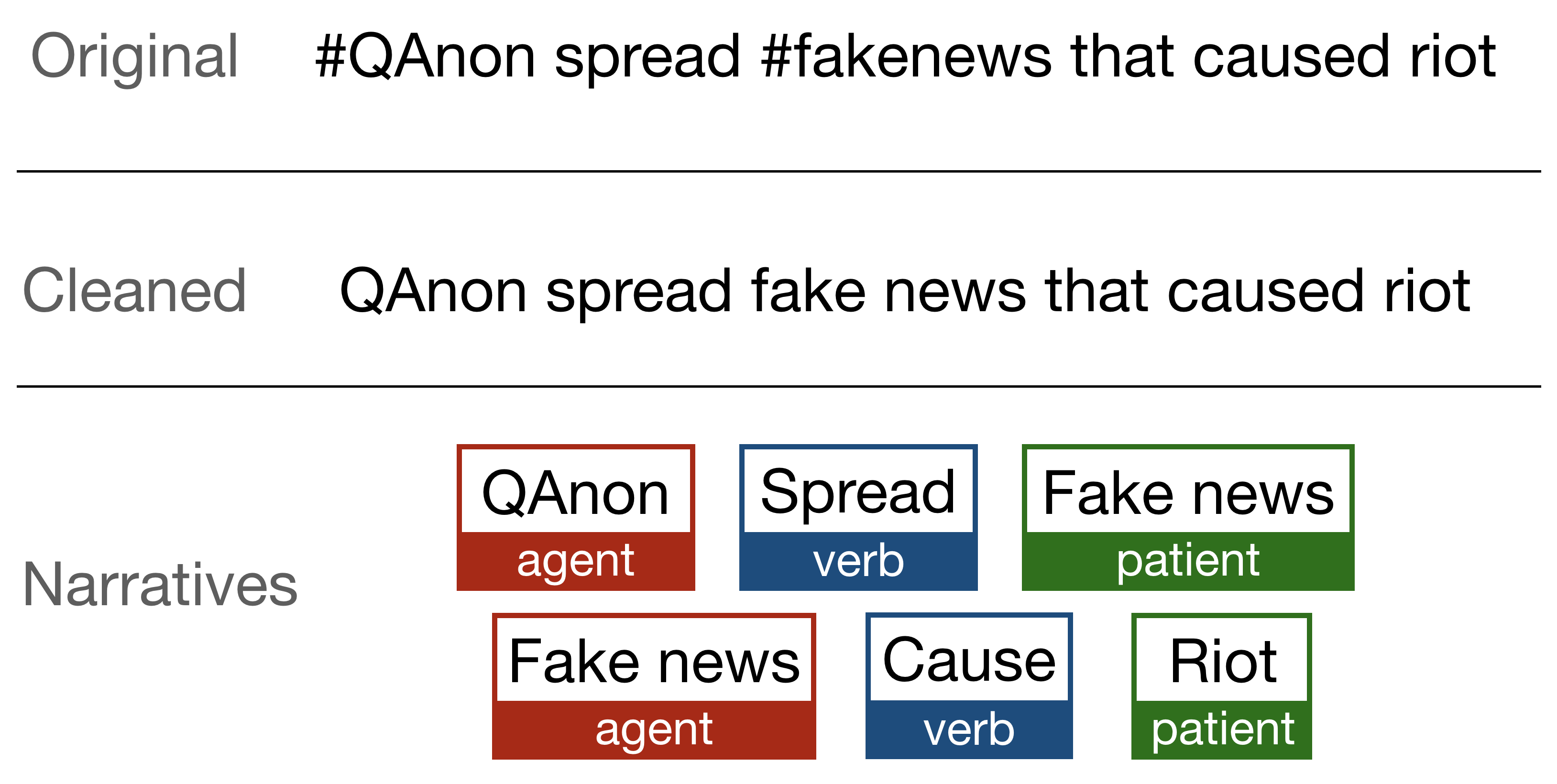}
    \caption{An example sentence being split into narratives}
    \label{fig:narrativeexample}
\end{figure}

While NER-based analysis allows us to measure the prevalence of terms on each platform, as well as undertake anti-social language analysis related to groups of interest, it does not offer insight into the \textit{context} in which those terms are mentioned. Simply knowing that something, for example QAnon, is mentioned a lot does not allow us to capture if one platform is overwhelmingly critical of QAnon, while the other is supportive. To deepen our understanding of content across the three platforms, we analyse narratives and inter-connected terms, using an adjusted Relatio library, and the method presented alongside it \cite{ash2021text}. We use Ash et al. operationalisation of narratives as triples of words or phrases that take the form \textit{Agent} - \textit{Verb} - \textit{Patient} (Figure \ref{fig:narrativeexample}).

\begin{figure}[t]
    \centering
    \includegraphics[width=\columnwidth]{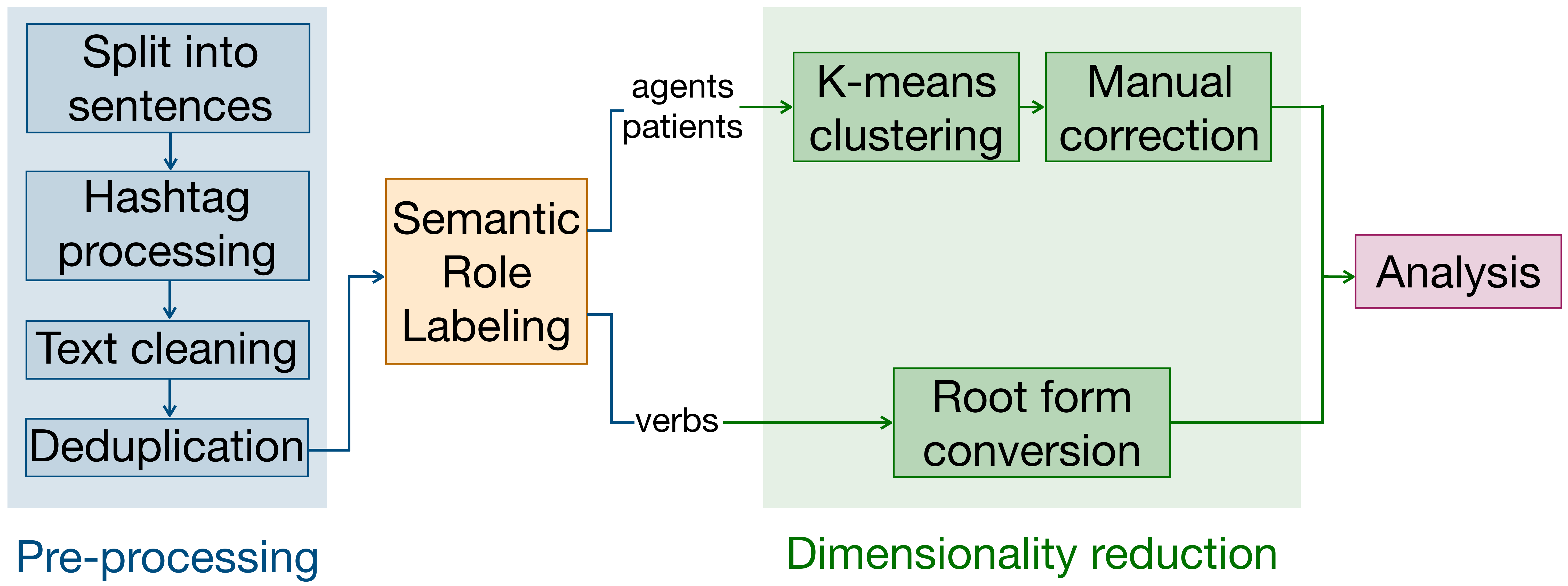}
    \caption{Narrative analysis pipeline}
    \label{fig:narrativepipeline}
\end{figure}

The narrative analysis pipeline is presented in Figure \ref{fig:narrativepipeline}. In pre-processing, we split posts into sentences, as the method can not reliably connect narratives spanning multiple sentences. Sentences can contain incomplete narratives (which we omit from analysis), or one or more complete narratives, such as the example we give in Figure \ref{fig:narrativeexample}. Special features of social media posts, such as hashtags, required adjustment of Ash et al. pre-processing. While some users use hashtags out of context, to boost visibility of their posts (in our case, especially on Parler as discussed previously), hashtags can be used in context, and simple removal would result in loss of information. To overcome this, we make best effort to remove hashtag blocks while preserving hashtags that are likely used in context. As some hashtags contain multiple words (e.g. \#fakenews in Figure \ref{fig:narrativeexample}), we manually establish how to split hashtags that occur more than five times on either platform. This maximises the information extracted from data, as we do not simply discard posts such as the one given in Figure \ref{fig:narrativeexample}. Finally, we clean text by removing mentions, links, emoji, and stopwords from posts, and remove duplicate posts made by the same person to ensure that spamming does not affect our analysis.

After pre-processing, we use semantic role labelling to identify roles in a sentence, extracting building blocks for narratives: verbs, agents and patients. This results in a vast vocabulary consisting of 100103 unique agents and patients, and 8213 unique verbs. We follow Ash et al.'s approach to reduce dimensionality by grouping agents and patients into similar themes with k-means clustering on GloVe embeddings. As our sample was not large enough to obtain a small number of high-quality clusters, we have manually corrected some of the categories (e.g. to ensure that mentions of Trump, Biden, and President are in separate clusters), resulting in 440 clusters representing agents and patients. We convert verbs to their root form, reducing the dimensionality of verbs to 3510. While dimensionality reduction results in some loss of information, it is necessary due to language diversity. A summary of the number of sentences and narratives is available in Table \ref{table:narrativesummary}. 

\begin{figure}[t]
    \centering
    \includegraphics[width=0.6\columnwidth]{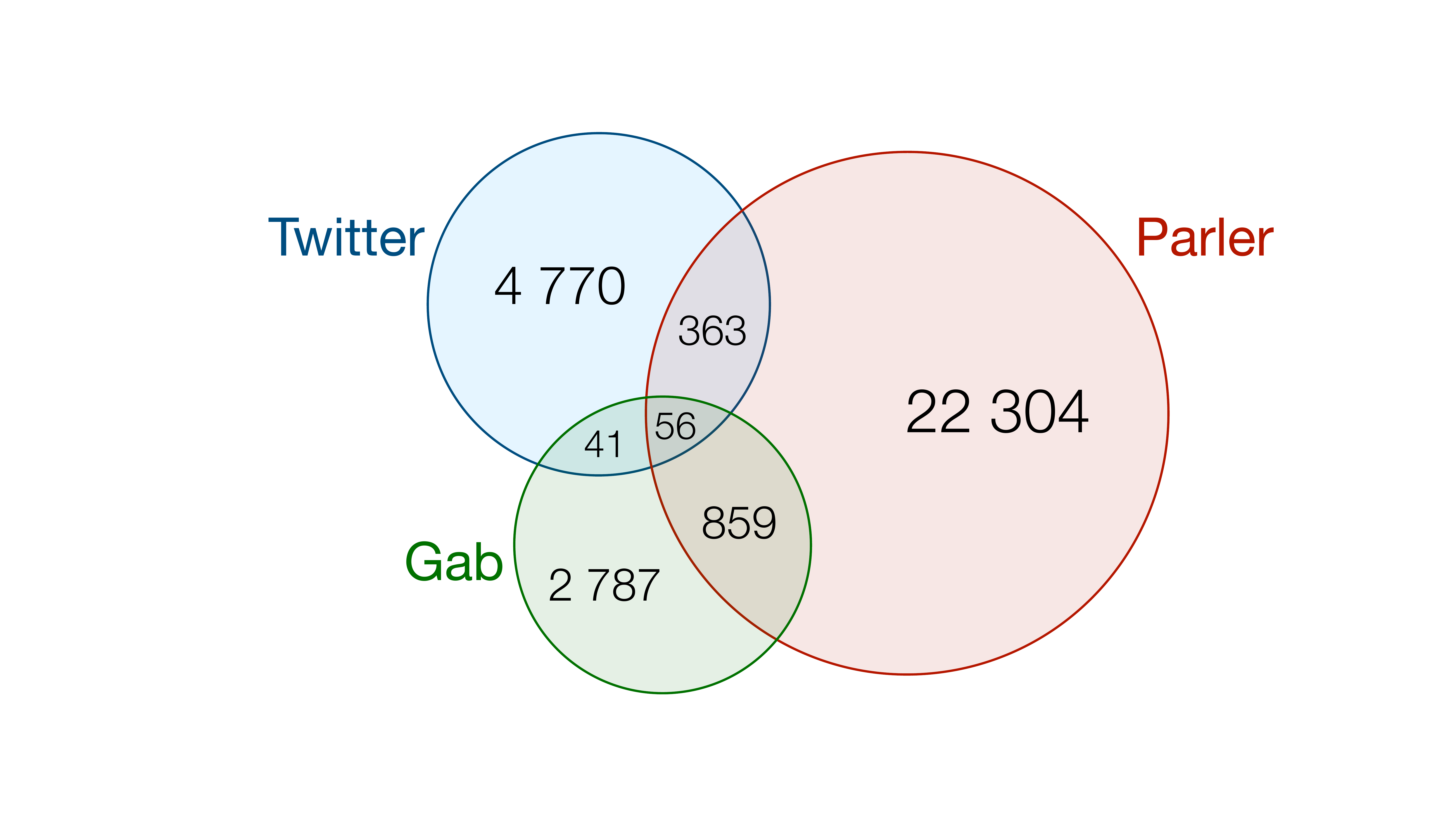}
    \caption{Overlap of unique narrative between platforms}
    \label{fig:narrativevenn}
\end{figure}

Of the 44395 complete narratives, 31180 are unique. The reason for this is a high number of unique verbs that connect Agents and Patients. While we reduced the dimensionality of verbs from 8213 to 3510, we could not reliably group them according to the similarity of action they represent. The overlap between unique narratives between the platforms is presented in Figure \ref{fig:narrativevenn}. 96\% of narratives appear on only one of the three platforms, with only 56 narratives appearing on all three. Parler shares more narratives with Gab than with Twitter, despite Gab having fewer unique narratives than Twitter. This indicates a higher similarity of the conversations happening on the Alt-Tech platforms compared to the mainstream platform.

\begin{figure*}
\subfloat[Twitter\label{fig:sfig1}]{\includegraphics[width= 0.32\textwidth]{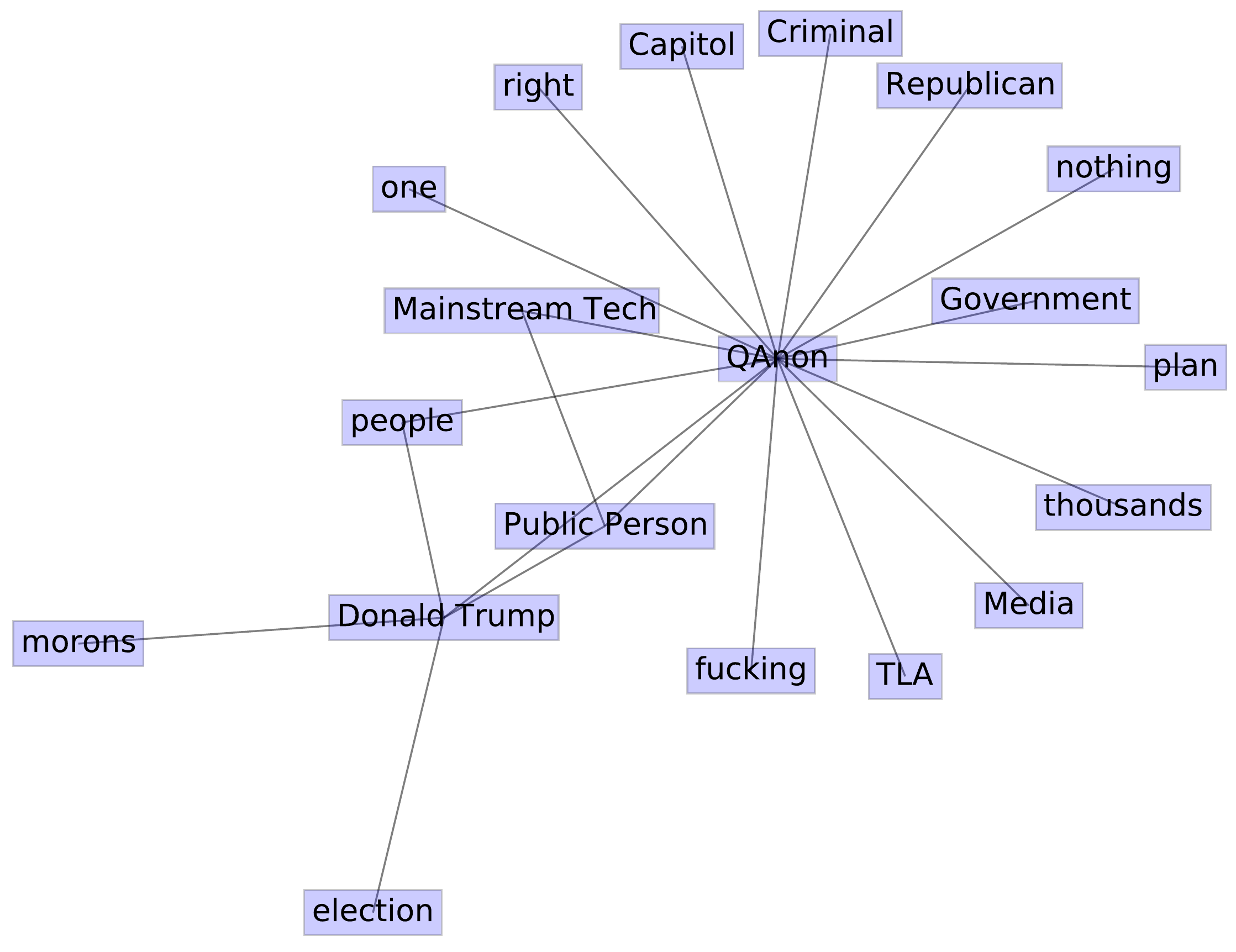}}
\subfloat[Parler\label{fig:sfig2}]{\includegraphics[width= 0.32\textwidth]{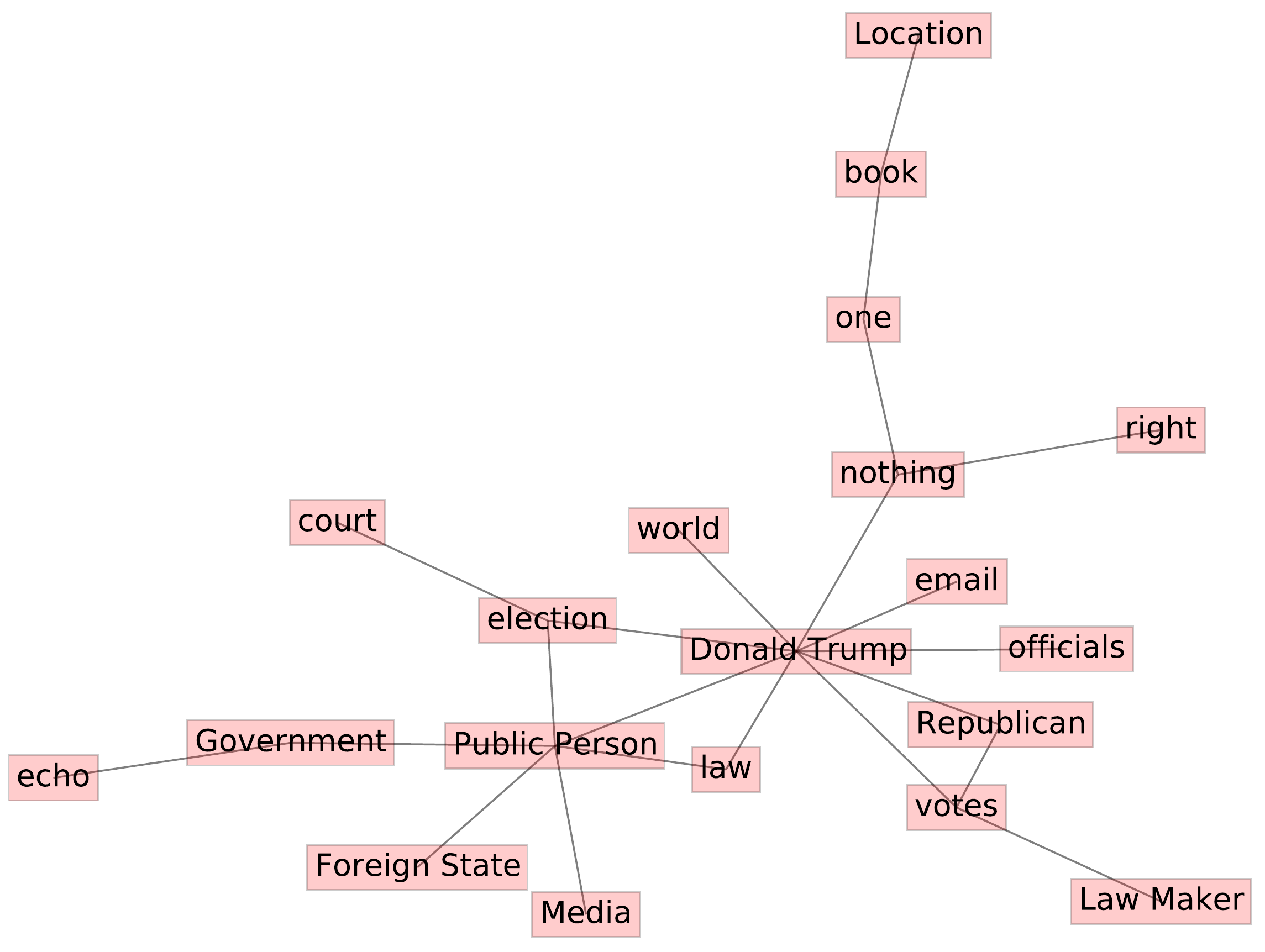}}
\subfloat[Gab\label{fig:sfig3}]{\includegraphics[width= 0.36\textwidth]{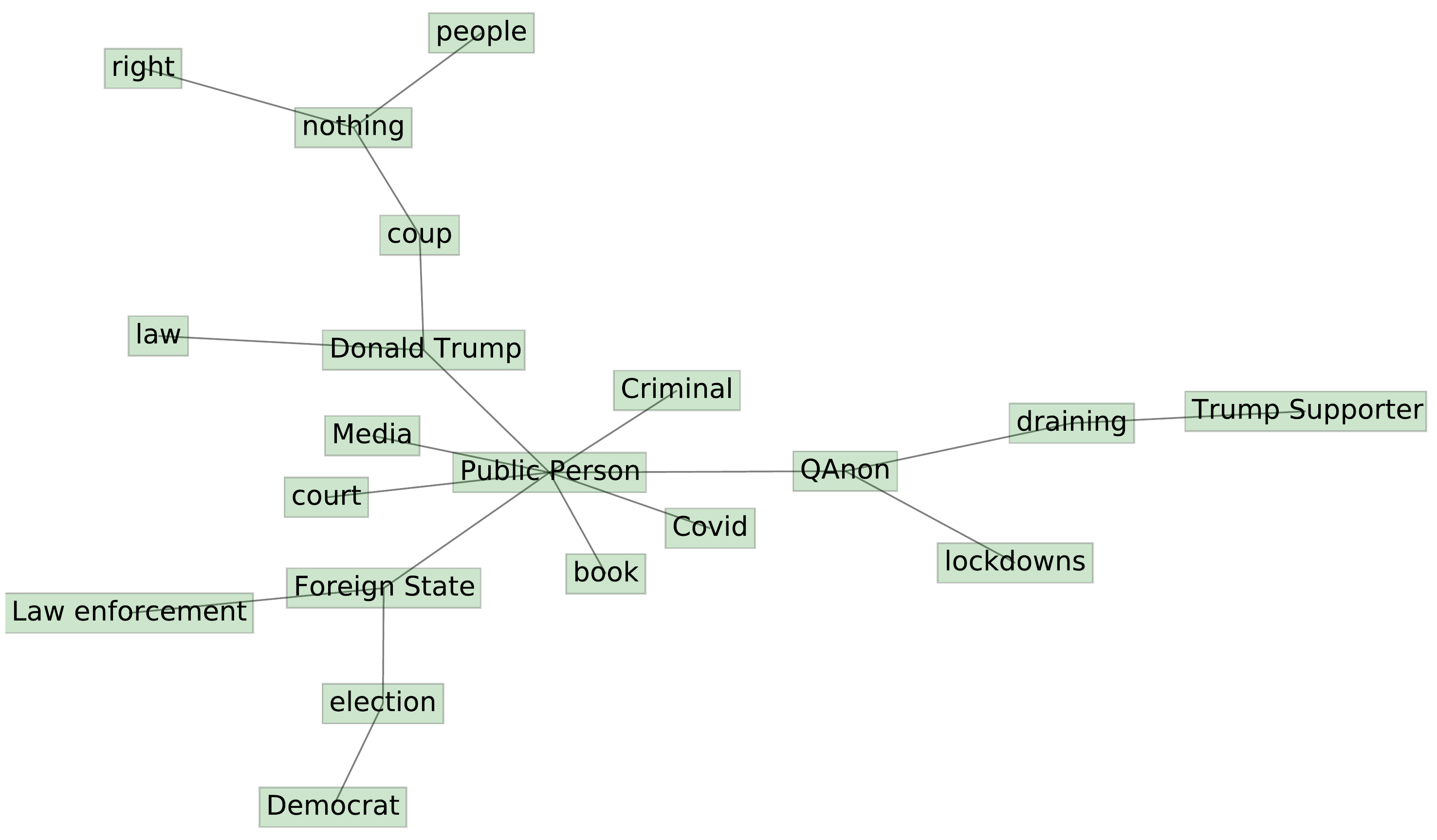}}
\caption{Giant connected components of the 30 most frequently co-occurring Agent and Patients}
\label{fig:noverb}
\end{figure*}

\begin{table}[t]
    \centering
    \resizebox{0.9\columnwidth}{!}{
    \begin{tabular}{llll}
        \toprule
         & Twitter & Parler & Gab \\
        \midrule
        Analysed posts          & 12759 &	81456   &	6997 \\
        Deduplicated posts      & 11412 &	55532   & 5076 \\
        Sentences               & 21306 &	111457  &	16122 \\
        Complete narratives     & 6629 & 32220 & 5546 \\
        Posts with complete narratives   & 4631  & 19656     & 2898 \\
        \bottomrule
    \end{tabular}}
    \caption{Summary on the number of narratives extracted}
    \label{table:narrativesummary}
\end{table}

The low overlap in narratives suggests that themes frequently appearing together differ substantially on the three platforms. To understand what are frequently co-occurring agents and patients, and how platforms differ in this regard, we model Agents and Patients as nodes in an undirected network. This network is agnostic to what verb connects the two but allows us to look at the most popular Agents and Patients on platforms, and which terms tend to appear together in a narrative. We present giant connected components for the 30 most frequent Agent-Patient pairs in Figure \ref{fig:noverb}. The three platforms differ substantially when it comes to what is central to the conversation - conversations on Twitter (Fig. \ref{fig:sfig1}) are highly centred on QAnon, which is expected given how we collected the data. Terms connected to QAnon represent political entities - such as media, government, Donald Trump, and public personalities (mainly overlapping with those observed in RQ3). In addition, we note the mention of Capitol, and that Twitter users linked it to QAnon. In stark contrast, QAnon does not even appear in the top 30 connections on Parler (Fig. \ref{fig:sfig2}). This is yet another indication of different use of hashtags on Parler, where \#QAnon was frequently used as part of hashtag blocks, rather than in the conversation context. Donald Trump and other public personalities take central roles of conversation on Parler, and the core is well-connected, except for ``Location'', ``Book'', and ``One'' (representing a person) making an appearance due to the high popularity of quote ``the man who reads nothing at all is better educated than the man who reads nothing but newspapers'' amongst Parler users. Gab network (Fig. \ref{fig:sfig3}) is more similar to Parler's, although public personalities are more connected than Trump himself, with QAnon making an appearance in relation to lockdowns, public personalities, and popular phrase used by Trump supporters ``Drain the swamp''.

Finally, we present the ten most prevalent narratives on all three platforms in Table \ref{table:topnarratives}. These results provide some context in which Agent-Patient pairs presented in Figure \ref{fig:noverb} appear together, although we note that even the most popular narratives are not used many times (some of the narratives in the top 10 appear as few as 5 times), a result of the high diversity of verbs used. QAnon is central to the discussion on Twitter, appearing in 6 top narratives, which are indicative of republicans' and Trump's relationship with QAnon, and the role that Twitter users believe QAnon had in the storming of the Capitol. On the other hand, Parler mostly sees the popularity of quotes (such as the example above) and calls to action by other Trump supporters - asking them to email him or follow and re-post. Gab has the highest level of conspiratorial ideas breaking into top 10 narratives - CIA infiltrating local states' governments, suggesting that nobody can prove QAnon is fake, and calling on Trump to invoke the Insurrection Act in response to the ``election being stolen''.

While the three platforms proved to be vastly similar when tested for anti-social language, prevalent topics of conversation, and which groups of political figures attract more anti-social language, narrative analysis reveals considerable differences in the context of mentions of these figures and how central QAnon itself is to the discussion. Twitter centres the conversation on QAnon, often mentioning it in a negative context; Parler focuses on expressing support for Trump and making calls to action towards like-minded users, while Gab narratives hint at a higher prominence and more central role of conspiratorial ideas.

\begin{table}[t]
    \centering
    \resizebox{\columnwidth}{!}{
    \begin{tabular}{lll}
        \toprule
        Twitter & Parler & Gab \\
        \midrule
        supporter show balls               & email send trump                  & trump supporter keep draining  \\
        republican push qanon               & person read book                  & qanon keep draining  \\
        qanon storm capitol                 & person read nothing               & nothing wake people  \\
        platform ban public person          & evil destroy world                & science use technology  \\
        trump delude morons                 & art support bot                   & trump invoke insurrection  \\
        qanon include legit                 & book honor location               & nothing stop coming \\
        trump meet qanon                    & government follow echo                 & activities affect military \\
        thousands believe qanon             & nothing stop right             & cia infiltrate us state \\
        qanon arrest crime                  & deep state engineer covid                & morons prove qanon \\
        trump supporter enter gov.          & evil destroy world         & lockdowns prove qanon \\
        \bottomrule
    \end{tabular}}
    \caption{10 most prevalent narratives}
    \label{table:topnarratives}
\end{table}

\section{Conclusions}
We have compared discourse around \#QAnon on three platforms - Twitter, Gab, and Parler - in a month preceding the storming of the US Capitol on January 6, 2021., using a variety of computational methods to capture different aspects of posts - from measures of anti-social language, to themes at the core of the discourse. Our findings show that the volumes of posting with this hashtag differ drastically across platforms, with Parler having the highest volume of data. While more unique users have posted using \#QAnon on Twitter, users on Parler and Gab made considerably more posts per user on average. We found that the prevalence of anti-social language on the three platforms varies depending on the measure. While Twitter and Parler emerge as leaders in terms of the distribution of posts with anti-social language based on the analysed Perspective API features, Gab has the highest proportion of posts with hate words.

The analysis of the most frequently mentioned named entities across the three platforms revealed important similarities and differences between them. On the one hand, there are obvious overlaps in the most popular named entities across the platforms, suggesting that \#QAnon-related discourse mentioned largely similar themes during the observation period. Such overlapping entities include Donald Trump, the US, and Washington, D.C. Entities enjoying higher popularity on Twitter compared to the other two platforms include individuals salient in the US political context in connection to QAnon (e.g., Marjorie Taylor Greene or Jake Angeli) as well as ``Antifa'' and ``Nazi'', suggesting divisive stance in relation to \#QAnon. On Gab and Parler, popular entities suggest a focus on discussions around the election results and related conspiracies (e.g., Dominion voting machines or Deep State) and the right to bear arms.

We observe differences in anti-social language measures in posts mentioning different political groups or individuals. Though the prevalence of such posts differs across platforms, as well as the prevalence of anti-social language in posts about different groups of politicians, we find that on all platforms posts mentioning female politicians, Democrats, and Donald Trump score higher on anti-social features than those with mentions of their male or Republican counterparts, or Joe Biden. Since our analysis focused on a very specific period and topic, it is unclear whether this observation can be generalised. We suggest that it is worthwhile to examine the prevalence of anti-social language in posts about different political groups in further cross-platform research.

Finally, our analysis of which terms appear together in posts, and the narratives they appear as part of, indicates that the core of discussions related to \#QAnon differs substantially across the three platforms. While Twitter focuses on QAnon and the most prevalent narratives related to it are critical of it, Parler focuses on supporting Donald Trump, while Gab sees a bigger focus on other political figures, as well as Trump, and has a higher prevalence of conspiratorial content amongst its most popular narratives.

Our study has limitations in addition to the ones listed in the Data section. Firstly, we focused on a single hashtag that, despite being one of the most popular in relation to QAnon conspiracy theory on all three platforms and one of the most popular hashtags in general on Parler, still reflects overall discourse surrounding QAnon only partially. Nonetheless, due to the prominence of this hashtag, we argue that the collected data reflects the most dominant aspects of QAnon-related discourse. Secondly, while the month-long time frame our study encompasses is appropriate for analysis of events that see a heightened volume of activity, a further study examining the development of QAnon-related discourse over time could be relevant, comparing the observations during the periods of high activity, as well as in more routine periods.

There are three implications of our findings that we wish to highlight in particular. Firstly, while the differences between the three platforms exist in our sample, they do not exactly align with what public opinion appears to be. Given the press Parler has received, and the consequences it has suffered in part related to the anti-social language, one might expect that Parler would exhibit a much higher prevalence of threatening, or toxic language than, for example, Twitter. Yet, our results show that platforms are largely comparable in this regard, with no clear ``villain'', and that platforms, including mainstream ones, can go further in limiting particularly harmful aspects of anti-social language. This also highlights the need for cross-platform comparative studies. The research community is uniquely equipped with methods, independence, and thoroughness required to make fair judgements to inform public perceptions, which might be currently largely based on preliminary analyses. Secondly, we show that more nuanced methods are necessary for comparative cross-platform studies, as single measures or simple metrics often do not reveal the differences, especially when observing multifaceted phenomena such as language and discourse. Finally, as we have seen with hashtags, seemingly identical functionalities can be used differently across different platforms, and pre-processing choices we make can substantially affect the conclusions of the analysis. This highlights the importance of the decisions we make at each step of our analyses, including pre-processing, as well as how transparently we communicate them in our studies, both in cross-platform research, and beyond.


\begin{thebibliography}{37}
\providecommand{\natexlab}[1]{#1}
\providecommand{\url}[1]{\texttt{#1}}
\providecommand{\urlprefix}{URL }
\expandafter\ifx\csname urlstyle\endcsname\relax
  \providecommand{\doi}[1]{doi:\discretionary{}{}{}#1}\else
  \providecommand{\doi}{doi:\discretionary{}{}{}\begingroup
  \urlstyle{rm}\Url}\fi

\bibitem[{Aliapoulios et~al.(2021{\natexlab{a}})Aliapoulios, Bevensee,
  Blackburn, Bradlyn, De~Cristofaro, Stringhini, and
  Zannettou}]{aliapoulios2021early}
Aliapoulios, M.; Bevensee, E.; Blackburn, J.; Bradlyn, B.; De~Cristofaro, E.;
  Stringhini, G.; and Zannettou, S. 2021{\natexlab{a}}.
\newblock An Early Look at the Parler Online Social Network.
\newblock \emph{arXiv preprint arXiv:2101.03820} .

\bibitem[{Aliapoulios et~al.(2021{\natexlab{b}})Aliapoulios, Papasavva,
  Ballard, De~Cristofaro, Stringhini, Zannettou, and
  Blackburn}]{aliapoulios2021gospel}
Aliapoulios, M.; Papasavva, A.; Ballard, C.; De~Cristofaro, E.; Stringhini, G.;
  Zannettou, S.; and Blackburn, J. 2021{\natexlab{b}}.
\newblock The gospel according to Q: Understanding the QAnon conspiracy from
  the perspective of canonical information.
\newblock \emph{arXiv preprint arXiv:2101.08750} .

\bibitem[{Ash, Gauthier, and Widmer(2021)}]{ash2021text}
Ash, E.; Gauthier, G.; and Widmer, P. 2021.
\newblock Text Semantics Capture Political and Economic Narratives.
\newblock \emph{arXiv preprint arXiv:2108.01720} .

\bibitem[{Barrie and Ho(2021)}]{academictwitteR}
Barrie, C.; and Ho, J. C.-t. 2021.
\newblock academictwitteR.
\newblock \urlprefix\url{https://github.com/cjbarrie/academictwitteR}.

\bibitem[{Brewster(2021)}]{brewster2021}
Brewster, T. 2021.
\newblock Sheryl Sandberg Downplayed Facebook’s Role In The Capitol Hill
  Siege—Justice Department Files Tell A Very Different Story.
\newblock \emph{Forbes} .

\bibitem[{{Castle Lemongrab}(2020)}]{parlance}
{Castle Lemongrab}. 2020.
\newblock Parlance API.
\newblock \url{https://github.com/castlelemongrab/parlance}.

\bibitem[{Chowdhury et~al.(2020)Chowdhury, Allen, Yousuf, and
  Mueen}]{chowdhury2020twitter}
Chowdhury, F.~A.; Allen, L.; Yousuf, M.; and Mueen, A. 2020.
\newblock On Twitter Purge: A Retrospective Analysis of Suspended Users.
\newblock In \emph{Companion Proceedings of the Web Conference 2020}.

\bibitem[{{EIP}(2020)}]{eip2020}
{EIP}. 2020.
\newblock Repeat Offenders: Voting Misinformation on Twitter in the 2020 United
  States Election.
\newblock Technical report, {Election Integrity Partnership}.

\bibitem[{ElSherief et~al.(2018)ElSherief, Kulkarni, Nguyen, Wang, and
  Belding}]{elsherief2018hate}
ElSherief, M.; Kulkarni, V.; Nguyen, D.; Wang, W.~Y.; and Belding, E. 2018.
\newblock Hate lingo: A target-based linguistic analysis of hate speech in
  social media.
\newblock In \emph{Proceedings of the International AAAI Conference on Web and
  Social Media}, volume~12.

\bibitem[{Ferrara et~al.(2020)Ferrara, Chang, Chen, Muric, and
  Patel}]{ferrara2020}
Ferrara, E.; Chang, H.; Chen, E.; Muric, G.; and Patel, J. 2020.
\newblock Characterizing social media manipulation in the 2020 U.S.
  presidential election.
\newblock \emph{First Monday} 25.

\bibitem[{Frenkel(2021)}]{frenkel2021}
Frenkel, S. 2021.
\newblock The storming of Capitol Hill was organized on social media.
\newblock \emph{The New York Times} .

\bibitem[{{Hatebase Inc.}(2021)}]{hatebase2021}
{Hatebase Inc.} 2021.
\newblock Hatebase API.
\newblock \url{https://hatebase.org}.

\bibitem[{Jigsaw(2021)}]{perspectiveFeatures}
Jigsaw. 2021.
\newblock Perspective API Attributes and languages.
\newblock
  \url{https://developers.perspectiveapi.com/s/about-the-api-attributes-and-languages}.

\bibitem[{Lima et~al.(2018)Lima, Reis, Melo, Murai, Araujo, Vikatos, and
  Benevenuto}]{lima2018inside}
Lima, L.; Reis, J.~C.; Melo, P.; Murai, F.; Araujo, L.; Vikatos, P.; and
  Benevenuto, F. 2018.
\newblock Inside the right-leaning echo chambers: Characterizing gab, an
  unmoderated social system.
\newblock In \emph{2018 IEEE/ACM International Conference on Advances in Social
  Networks Analysis and Mining}.

\bibitem[{McIlroy-Young and Anderson(2019)}]{mcilroy2019welcome}
McIlroy-Young, R.; and Anderson, A. 2019.
\newblock From “welcome new gabbers” to the pittsburgh synagogue shooting:
  The evolution of gab.
\newblock In \emph{Proceedings of the International AAAI Conference on Web and
  Social Media}, volume~13.

\bibitem[{Olteanu et~al.(2018)Olteanu, Castillo, Boy, and
  Varshney}]{olteanu2018effect}
Olteanu, A.; Castillo, C.; Boy, J.; and Varshney, K. 2018.
\newblock The effect of extremist violence on hateful speech online.
\newblock In \emph{Proceedings of the International AAAI Conference on Web and
  Social Media}, volume~12.

\bibitem[{Papakyriakopoulos, Serrano, and
  Hegelich(2020)}]{papakyriakopoulos2020spread}
Papakyriakopoulos, O.; Serrano, J. C.~M.; and Hegelich, S. 2020.
\newblock The spread of COVID-19 conspiracy theories on social media and the
  effect of content moderation.
\newblock \emph{Harvard Kennedy School Misinformation Review} 10.

\bibitem[{Pavlopoulos et~al.(2020)Pavlopoulos, Sorensen, Dixon, Thain, and
  Androutsopoulos}]{pavlopoulos2020toxicity}
Pavlopoulos, J.; Sorensen, J.; Dixon, L.; Thain, N.; and Androutsopoulos, I.
  2020.
\newblock Toxicity Detection: Does Context Really Matter?
\newblock In \emph{Proceedings of the 58th Annual Meeting of the Association
  for Computational Linguistics}.

\bibitem[{Qi et~al.(2020)Qi, Zhang, Zhang, Bolton, and Manning}]{qi2020stanza}
Qi, P.; Zhang, Y.; Zhang, Y.; Bolton, J.; and Manning, C.~D. 2020.
\newblock Stanza: A {Python} Natural Language Processing Toolkit for Many Human
  Languages.
\newblock In \emph{Proceedings of the 58th Annual Meeting of the Association
  for Computational Linguistics: System Demonstrations}.

\bibitem[{Rheault, Rayment, and Musulan(2019)}]{rheault2019politicians}
Rheault, L.; Rayment, E.; and Musulan, A. 2019.
\newblock Politicians in the line of fire: Incivility and the treatment of
  women on social media.
\newblock \emph{Research \& Politics} 6(1).

\bibitem[{Sap et~al.(2019)Sap, Card, Gabriel, Choi, and Smith}]{sap2019risk}
Sap, M.; Card, D.; Gabriel, S.; Choi, Y.; and Smith, N.~A. 2019.
\newblock The risk of racial bias in hate speech detection.
\newblock In \emph{Proceedings of the 57th annual meeting of the association
  for computational linguistics}.

\bibitem[{Siegel et~al.(2019)Siegel, Nikitin, Sterling, Pullen, Bonneau,
  Nagler, Tucker et~al.}]{siegel16trumping}
Siegel, A.~A.; Nikitin, E.; Sterling, J.; Pullen, B.; Bonneau, R.; Nagler, J.;
  Tucker, J.~A.; et~al. 2019.
\newblock Trumping Hate on Twitter? Online Hate Speech in the 2016 US Election
  Campaign and its Aftermath.
\newblock \emph{Quarterly Journal of Political Science} 16(1).

\bibitem[{Silva et~al.(2016)Silva, Mondal, Correa, Benevenuto, and
  Weber}]{silva2016analyzing}
Silva, L.; Mondal, M.; Correa, D.; Benevenuto, F.; and Weber, I. 2016.
\newblock Analyzing the targets of hate in online social media.
\newblock In \emph{Proceedings of the International AAAI Conference on Web and
  Social Media}, volume~10.

\bibitem[{Stevens(2020)}]{garc}
Stevens, C. 2020.
\newblock garc: Python and Command-Line Interface for Gab.com API.
\newblock \url{https://github.com/ChrisStevens/garc}.

\bibitem[{Tahmasbi et~al.(2021)Tahmasbi, Schild, Ling, Blackburn, Stringhini,
  Zhang, and Zannettou}]{tahmasbi2021go}
Tahmasbi, F.; Schild, L.; Ling, C.; Blackburn, J.; Stringhini, G.; Zhang, Y.;
  and Zannettou, S. 2021.
\newblock “Go eat a bat, Chang!”: On the Emergence of Sinophobic Behavior
  on Web Communities in the Face of COVID-19.
\newblock In \emph{Proceedings of the Web Conference 2021}, 1122--1133.

\bibitem[{Thiel et~al.(2021)Thiel, DiResta, Grossman, and
  Cryst}]{thiel2021contours}
Thiel, D.; DiResta, R.; Grossman, S.; and Cryst, E. 2021.
\newblock Contours and Controversies of Parler.
\newblock Technical report, Stanford University.

\bibitem[{Timberg and Harwell(2021)}]{timberg2021}
Timberg, C.; and Harwell, D. 2021.
\newblock Pro-Trump forums erupt with violent threats ahead of Wednesday’s
  rally against the 2020 election.
\newblock \emph{The Washington Post} .

\bibitem[{Twitter(2021)}]{twitterAcademic2021}
Twitter. 2021.
\newblock Twitter Academic Research API.
\newblock
  \urlprefix\url{https://developer.twitter.com/en/solutions/academic-research/}.

\bibitem[{TwitterSafety(2020)}]{twitter2020}
TwitterSafety. 2020.
\newblock Twitter's announcement regarding moderation of QAnon.
\newblock \url{https://twitter.com/TwitterSafety/status/1285726277719199746}.

\bibitem[{{US Department of Justice}(2021)}]{angeliCharge}
{US Department of Justice}. 2021.
\newblock Three Men Charged in Connection with Events at U.S. Capitol.
\newblock
  \urlprefix\url{https://www.justice.gov/usao-dc/pr/three-men-charged-connection-events-us-capitol}.

\bibitem[{Wang et~al.(2021)Wang, Zannettou, Blackburn, Bradlyn, De~Cristofaro,
  and Stringhini}]{wang2021multi}
Wang, Y.; Zannettou, S.; Blackburn, J.; Bradlyn, B.; De~Cristofaro, E.; and
  Stringhini, G. 2021.
\newblock A Multi-Platform Analysis of Political News Discussion and Sharing on
  Web Communities.
\newblock \emph{arXiv preprint arXiv:2103.03631} .

\bibitem[{Wilson(2021)}]{wilson2021}
Wilson, J. 2021.
\newblock Rightwingers flock to 'alt tech' networks as mainstream sites ban
  Trump.
\newblock \emph{The Guardian} .

\bibitem[{Zannettou et~al.(2018)Zannettou, Bradlyn, De~Cristofaro, Kwak,
  Sirivianos, Stringini, and Blackburn}]{zannettou2018gab}
Zannettou, S.; Bradlyn, B.; De~Cristofaro, E.; Kwak, H.; Sirivianos, M.;
  Stringini, G.; and Blackburn, J. 2018.
\newblock What is gab: A bastion of free speech or an alt-right echo chamber.
\newblock In \emph{Companion Proceedings of the The Web Conference 2018}.

\bibitem[{Zannettou et~al.(2020{\natexlab{a}})Zannettou, ElSherief, Belding,
  Nilizadeh, and Stringhini}]{zannettou2020measuring}
Zannettou, S.; ElSherief, M.; Belding, E.; Nilizadeh, S.; and Stringhini, G.
  2020{\natexlab{a}}.
\newblock Measuring and characterizing hate speech on news websites.
\newblock In \emph{12th ACM Conference on Web Science}.

\bibitem[{Zannettou et~al.(2020{\natexlab{b}})Zannettou, Finkelstein, Bradlyn,
  and Blackburn}]{zannettou2020quantitative}
Zannettou, S.; Finkelstein, J.; Bradlyn, B.; and Blackburn, J.
  2020{\natexlab{b}}.
\newblock A quantitative approach to understanding online antisemitism.
\newblock In \emph{Proceedings of the International AAAI Conference on Web and
  Social Media}, volume~14.

\bibitem[{Zeng and Sch{\"a}fer(2021)}]{zeng2021conceptualizing}
Zeng, J.; and Sch{\"a}fer, M.~S. 2021.
\newblock Conceptualizing “Dark Platforms”. Covid-19-Related Conspiracy
  Theories on 8kun and Gab.
\newblock \emph{Digital Journalism} 1--23.

\bibitem[{Zuckerman(2019)}]{Zuckerman2019QAnon}
Zuckerman, E. 2019.
\newblock QAnon and the Emergence of the Unreal.
\newblock \emph{Journal of Design and Science} 1(6).

\end{thebibliography}
\end{document}